\documentstyle[psfig,times]{mn}

\def\etal{et~al.}
\def\spose#1{\hbox to 0pt{#1\hss}}
\def\lta{\mathrel{\spose{\lower 3pt\hbox{$\mathchar"218$}}
     \raise 2.0pt\hbox{$\mathchar"13C$}}}
\def\gta{\mathrel{\spose{\lower 3pt\hbox{$\mathchar"218$}}
     \raise 2.0pt\hbox{$\mathchar"13E$}}}

\def\clean{{\sc clean}}

\def\uv{{\it uv}}

\def\aips{{\sc aips}}

\def\fits{{\sc fits}}
\def\sad{{\sc sad}}

\def\imagr{{\sc imagr}}

\title[CENSORS: A Combined EIS-NVSS Survey Of Radio Sources]{CENSORS: A
Combined EIS-NVSS Survey Of Radio Sources.\\ 
I. Sample definition, radio data and optical identifications}

\author[P.~N.~Best \etal]{P.~N.~Best,$^1$\thanks{Email:
pnb@roe.ac.uk} J.~N.~Arts$^2$, H.~J.~A.~R{\"o}ttgering$^2$, R.~Rengelink$^2$,
M.~H.~Brookes$^1$, J.~Wall$^3$ \\ 
$^1$ Institute for Astronomy, Royal Observatory Edinburgh, Blackford Hill,
Edinburgh EH9 3HJ, UK\\
$^2$ Sterrewacht Leiden, Postbus 9513, 2300 RA Leiden, the Netherlands\\ 
$^3$ Department of Astrophysics, University of Oxford, Keble Road, Oxford,
OX1 3RH, United Kingdom
}

\pagerange{\pageref{firstpage}--\pageref{lastpage}}
\pubyear{2002}
\begin{document}
\label{firstpage}

\maketitle

\begin{abstract}
\noindent 
A new sample of radio sources, with the designated name CENSORS ({\it A
Combined EIS-NVSS Survey Of Radio Sources}), has been defined by combining the
NRAO VLA Sky Survey (NVSS) at 1.4\,GHz with the ESO Imaging Survey (EIS) Patch
D, a 3 by 2 degree region of sky centred at 09 51 36.0, $-$21 00 00 (J2000).
New radio observations of 199 NVSS radio sources with NVSS flux densities
$S_{1.4\,GHz} >7.8$\,mJy are presented, and compared with the EIS $I$--band
imaging observations which reach a depth of $I \sim 23$; optical
identifications are obtained for over two--thirds of the $\sim 150$ confirmed
radio sources within the EIS field. The radio sources have a median linear
size of 6 arcseconds, consistent with the trend for lower flux density radio
sources to be less extended. Other radio source properties, such as the lobe
flux density ratios, are consistent with those of brighter radio source
samples. From the optical information, 30--40\% of the sources are expected to
lie at redshifts $z \gta 1.5$.

One of the key goals of this survey is to accurately determine the high
redshift evolution of the radio luminosity function.  These radio sources are
at the ideal flux density level to achieve this goal; at redshifts $z \sim 2$
they have luminosities which are around the break of the luminosity function
and so provide a much more accurate census of the radio source population at
those redshifts than the existing studies of extreme, high radio power
sources.  Other survey goals include investigating the dual--population
unification schemes for radio sources, studying the radio luminosity
dependence of the evolution of radio source environments, and understanding
the radio power dependence of the K$-z$ relation for radio galaxies.
\end{abstract}


\section{Introduction}

The study of complete samples of radio sources can provide important
information on the cosmic evolution of the number density, nature and physical
properties of radio sources. For this reason, since the birth of radio
astronomy considerable effort has been put into obtaining complete optical
identifications and redshifts for flux--limited samples of sources selected at
a variety of different observing frequencies and flux density levels (e.g.
Laing, Riley \& Longair 1983 Allington-Smith 1984, Spinrad \etal\ 1985, Eales
1985a, Wall \& Peacock 1985, Dunlop \etal\ 1989, Lacy \etal\ 1993, McCarthy
\etal\ 1996, Best, R\"ottgering \& Lehnert 1999,2000, Waddington \etal\ 2001,
Rawlings, Eales \& Lacy 2001, Willott \etal\ 2002, and references therein).
\nocite{wal85,lai83,all84,eal85c,dun89b,wil02,raw01,wad01,lac93,bes99e}
\nocite{bes00a,spi85a,mcc96b}

An important study that can be carried out with such samples is to determine
the cosmic evolution of the luminosity function of radio sources.  In 1990,
Dunlop \& Peacock \shortcite{dun90} carried out a detailed investigation of
this, and presented the first evidence for a decline in the comoving number
density of powerful radio sources beyond $z \sim 2.5$. Since this time,
however, whilst numerous advances have provided a good consensus in the
determination of the low redshift radio luminosity function (e.g. Mobasher
\etal\ 1999, Gavazzi \& Boselli 1999, Machalski \& Godlowski 2000, Sadler
\etal\ 2002)\nocite{mob99,gav99,mac00,sad02}, the high redshift evolution and
the reality of the `redshift cut-off' in the radio source population have
remained areas of much controversy. Willott \etal\ \shortcite{wil98,wil01}
found no evidence for a cut-off beyond $z \sim 2$ using the 7C radio survey
\cite{poo98}, whilst Bremer \etal\ \shortcite{bre99b} claimed tentative
indications of a redshift cut-off in a smaller sample of ultra-steep spectrum
sources selected from the Westerbork Northern Sky Survey (WENSS; Rengelink
\etal\ 1998)\nocite{ren98} at similar luminosities. In more recent studies,
Waddington \etal\ \shortcite{wad01} showed that a sample selected at the
1\,mJy level does show evidence for a deficit of moderate luminosity radio
sources at $z > 2$, but this sample had insufficient sky coverage to
investigate the most luminous sources, while the sample of Jarvis \etal\
\shortcite{jar01} at the 100 mJy level proved too shallow to detect sufficient
sources at high enough redshift to clarify the high-redshift evolution. It is
clear that a radio source sample at a flux density level intermediate between
these two is required to resolve this issue.

Taking a census of the high redshift space density of radio sources is one of
the key goals of the current survey. In recent years it has become apparent
that resolving this issue has far-reaching importance, since massive black
holes appear to reside in all massive present-day spheroids, with a mass
roughly proportional to the baryonic mass of the spheroid \cite{kor01}. This
suggests that black-hole and spheroid formation are intimately linked
(e.g. Richstone \etal\ 1998)\nocite{ric98b} and that understanding the cosmic
evolution of black holes is of importance for testing theories of structure
formation in general. There are strong indications that powerful radio
activity, at least for steep spectrum radio sources (see Woo \& Urry 2002 for
an alternative view based on a study of flat--spectrum
quasars)\nocite{woo02}, is only produced by the most massive black holes
($M \gta 10^9 M_{\odot}$; e.g. Dunlop \etal\ 2001); the cosmic evolution of
powerful radio sources may therefore offer the cleanest way to constrain the
evolution of the top end of the black-hole mass function.

A complete radio source sample at these flux densities will also enable
several other important astrophysical questions to be addressed. These
include:

\begin{itemize}
\item Dual--population unification schemes of radio sources: Wall \& Jackson
\shortcite{wal97} model the radio luminosity function using just 2 parent
populations of radio sources, Fanaroff \& Riley (1974; hereafter
FR)\nocite{fan74} Class I and II, in conjunction with orientation unification.
The population mix and evolution at the low luminosities of this sample will
provide a critical test of this model, as well as allowing investigation of
the differential evolution of FR\,I and FR\,II radio sources (c.f. Snellen \&
Best 2001).\nocite{sne01}

\item Evolution of radio source environments: studies of the environments
around radio sources of different radio luminosities (e.g. Wold \etal\ 2000,
Finn \etal\ 2001)\nocite{wol00,fin01} show that at moderate redshifts ($z \sim
0.5$) the most luminous radio sources lie in richer environments than lower
luminosity sources, in contrast to what is found in the nearby Universe
(e.g. Prestage \& Peacock 1988)\nocite{pre88}.  The inconsistency of the
spatial correlation functions of the Green Bank and WENSS surveys
\cite{ren98}, and the weak variation of the cross--correlation amplitude of
radio sources with radio flux density (e.g. Overzier \etal\ 2003 and
references therein)\nocite{ove03} further suggest that the evolution of radio
source environments may be radio luminosity dependent, in a similar manner to
evolution of the comoving number densities of sources.  A sample of lower
luminosity sources would test this model providing valuable insight into the
physical mechanisms behind radio source evolution.

\item The K$-z$ relation: radio galaxies show a very tight correlation between
their K-magnitude and redshift \cite{lil84a}, but interestingly, the K$-z$
relations of radio galaxies from the 3CR and 6C samples, selected at different
limiting radio flux densities, are in agreement at low redshifts ($z \lta
0.6$) but show a mean offset of $\sim 0.6$ magnitudes in the K-band at $z \gta
1$ \cite{eal96,ins02a}. This result means either that the K$-$band magnitudes
of the most radio luminous sources contain a significant AGN contribution,
which would have important consequences for interpretations of the spectral
energy distributions of radio galaxies in terms of their stellar populations,
or that more powerful radio sources at high redshift are hosted by more
massive galaxies \cite{bes98d}. This second possibility is consistent with the
black--hole vs spheroid mass correlation, since if the black holes in high
redshift radio galaxies are fuelled at the Eddington limit then more massive
galaxies (with more massive black holes) will produce more powerful radio
sources. The similarity of the K-magnitudes of the different radio samples at
lower redshifts would then have interesting implications for the evolution of
fuelling of these objects. Comparing the K$-z$ relation for fainter radio
samples with that of the 6C sample will help to distinguish between these two
possibilities (c.f. also Willott \etal\ 2003).\nocite{wil03}
\end{itemize}

With all of these goals in mind, we have begun a project to produce a
spectroscopically complete sample of $\sim 150$ radio sources at the $S_{\rm
1.4\,GHz} \sim 10$\,mJy level. This project takes advantage of two recent
large surveys: the NRAO VLA Sky Survey (NVSS) which has surveyed the radio sky
at 1.4\,GHz, and the ESO Imaging Survey (EIS) which has provided deep optical
imaging over four 3 by 2 degree fields. By combining these two surveys, a
sample of radio sources can be defined for which the majority already have
optical identifications for their host galaxies.

In this paper, the radio source sample is defined, the radio properties of the
sample are described based on new high angular resolution observations, and a
comparison of these with the EIS optical data is made. The layout of the paper
is as follows. Section~\ref{sampreq} discusses the properties (e.g. depth,
size) required of a radio source sample to successfully address the question
of the high redshift evolution of the radio luminosity function. In
Section~\ref{sampdef} the EIS and NVSS surveys are briefly described, together
with the definition of the preliminary EIS-NVSS sample from these. The new
higher resolution VLA radio observations of these sources are described in
Section~\ref{obssect}, and the radio properties of these sources are discussed
in Section~\ref{radprops}.  The new radio maps are compared with the EIS
optical fields in Section~\ref{optids}, together with a likelihood analysis to
identify the possible optical host galaxies; the nature of these optical hosts
is discussed. The new data are used to refine the sample definition and
produce the final radio source sample, designated the CENSORS sample (a {\it
Combined EIS--NVSS Survey of Radio Sources}), in Section~\ref{finalsamp}. The
results are summarised in Section~\ref{concs}.  Subsequent papers (Brookes
\etal\ in prep.)  will present infrared K--band imaging data to identify the
radio source host galaxies too faint to appear in the EIS survey, the
spectroscopic follow--up observations, and the consequences of these for our
understanding of the evolution of the radio luminosity function at high
redshifts.

\section{Sample Requirements}
\label{sampreq}

From the results of Waddington \etal\ \shortcite{wad01} and Jarvis \etal\
\shortcite{jar01} it is clear that to resolve the question of the redshift
cut--off, what is required is near-complete redshift information for a sample
selected at intermediate flux densities, $S \sim 10$ mJy. This corresponds to
about the break of the radio luminosity function at redshifts $z \sim 2$;
10\,mJy radio sources at redshifts $z \gta 2$ are also of comparable radio
luminosity to nearby luminous radio sources such as those in the 3CR sample,
allowing the most direct comparison of the cosmic evolution of the intrinsic
properties of the radio sources.

\begin{figure}
\begin{tabular}{cc}
\psfig{file=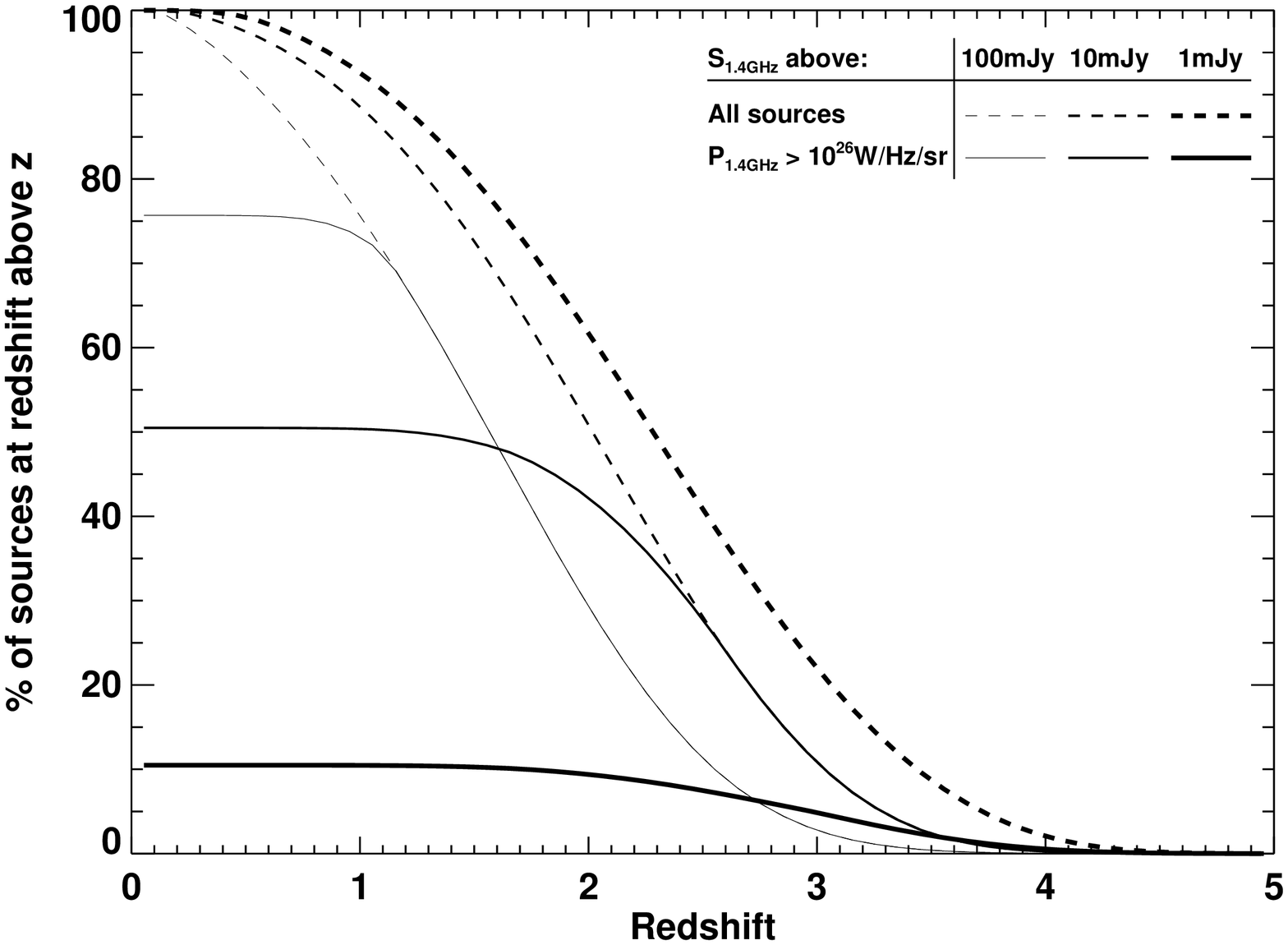,angle=90,width=8cm,clip=}\\
\psfig{file=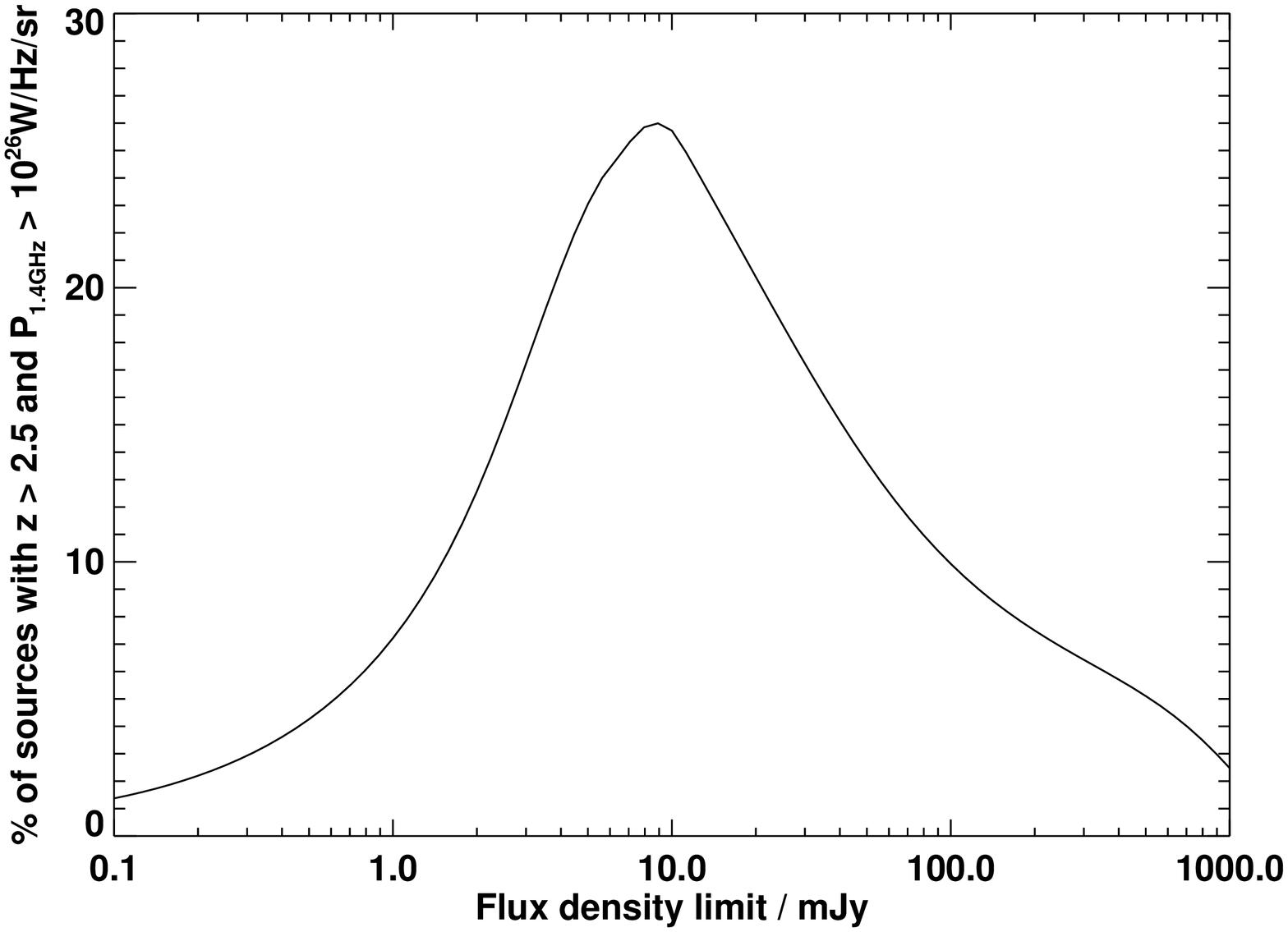,angle=90,width=8cm,clip=}
\end{tabular}
\caption{\label{dunpeacfig} {\it Top:} The dotted lines show the percentage of
sources in a flux-limited radio source sample which lie above a given
redshift, for three different flux density limits (100\,mJy, 10\,mJy and
1\,mJy, in order of increasing line thickness). The solid lines indicate the
subset of those sources which have a radio power $P_{\rm 1.4GHz} >
10^{26}$W\,Hz$^{-1}$\,sr$^{-1}$, which corresponds roughly to the break in
luminosity function. {\it Bottom:} The percentage of sources in a flux-limited
radio source sample which have $z>2.5$ and $P_{\rm 1.4GHz} >
10^{26}$W\,Hz$^{-1}$\,sr$^{-1}$, as a function of the flux density limit of
the sample. All of these results are based upon the `pure luminosity
evolution' model of Dunlop \& Peacock (1990), for steep-spectrum sources only,
assuming no redshift cut-off. The flux densities have been converted from 2.7
to 1.4\,GHz assuming a spectral index of 0.75. The Dunlop \& Peacock models,
which had been derived for an Einstein-de-Sitter cosmology, were converted to
the currently favoured lambda cosmology ($\Omega_m = 0.3$, $\Omega_{\Lambda} =
0.7$, $H_0 = 65$\,km\,s$^{-1}$Mpc$^{-1}$) using the expression provided by
Peacock (1985): $\rho_1(L_1,z) {\rm d}V_1/{\rm d}z = \rho_2(L_2,z) {\rm
d}V_2/{\rm d}z$, where $L_1$ and $L_2$ are the luminosities derived for a
source of given flux density and redshift in the two cosmologies, $V_1$ and
$V_2$ are the volumes available at a given redshift, and $\rho_1$ and $\rho_2$
are the corresponding space densities of sources. Willott \etal\ (2001) showed
that this conversion is accurate for all regions where the data constrain the
models well, although it is poorer in less-constrained regions of parameter
space. The use of different evolution models will also change the precise
details of these plots, but the general result that a 10\,mJy flux density
limit is optimal for these studies should be robust.}
\end{figure}
\nocite{pea85,wil01}

The suitability of the 10\,mJy flux density level for radio luminosity
function work is demonstrated in Figure~\ref{dunpeacfig}, where the percentage
of the radio sources in a sample above a given radio power and redshift is
illustrated as a function of redshift for three different flux density limits
(1, 10 and 100\,mJy at 1.4\,GHz), for the `Pure luminosity evolution' model of
Dunlop \& Peacock \shortcite{dun90} with no redshift cut-off. It is clear that
a 10\,mJy flux density cut-off provides by far the largest fraction of
powerful sources at redshifts beyond $z \sim 2$. The lower panel of
Figure~\ref{dunpeacfig} demonstrates this further, by comparing the percentage
of sources in a sample which have radio powers $P_{\rm 1.4GHz} >
10^{26}$W\,Hz$^{-1}$sr$^{-1}$ (around the break of the radio luminosity
function at $z \sim 2$) and redshifts $z > 2.5$ as a function of the flux
density limit of the sample. Although the exact details are model dependent, a
10\,mJy sample is clearly at around the optimal flux density level to
distinguish the presence or absence of a redshift cut--off.

Another important issue, given the goal of measuring the space density of high
redshift radio sources, is that the area of sky studied is large enough not to
be significantly affected by large--scale structure, especially since these
radio galaxies are likely to be highly clustered sources. 

Clustering increases the chances of finding a galaxy at a distance $r$ from
another galaxy by a factor $\xi(r)$, such that the probability of finding two
galaxies in two volumes $dV_1$ and $dV_2$ separated by a distance $r$ is given
by $P(r) = N^2 [1+\xi(r)] dV_1 dV_2$, where $N$ is the number of galaxies per
unit volume. The cross-correlation function $\xi(r)$ is known to be
well--matched by a power--law, $\xi(r) = (r/r_0)^{-\gamma}$, where $r_0$ is
the correlation length and $\gamma \approx 1.8$. Galaxies at redshifts $z \sim
3$ selected by the Lyman-break technique have been shown to have correlation
lengths of $r_0 \approx 3-4$\,Mpc \cite{por02}. Radio galaxy hosts are likely
to be more massive than these, and hence more strongly clustered; Daddi \etal\
\shortcite{dad03} find that luminous red (J$-$K $>1.7$) galaxies with
photometric redshifts $2 \lta z_{\rm phot} \lta 4$ have correlation lengths of
$\sim 10-12$\,Mpc, and these are more likely to be representative of the radio
source population. At redshifts $z \sim 3$, a spatial scale of 10\,Mpc
corresponds to about 20 arcminutes on the sky ($\Omega_m = 0.3$,
$\Omega_{\Lambda} = 0.7$, $H_0 = 65$\,km\,s$^{-1}$Mpc$^{-1}$).

To assess the affect of this clustering on the high redshift radio source
counts, a population of radio sources was built up, following the method
outlined by Soneira \& Peebles \shortcite{son77,son78}, such that they had a
correlation length of 12\,Mpc. These were constructed over a $2000 \times 2000
\times 2000$\,Mpc$^3$ volume. A rectangular field--of--view was then chosen,
of a given long axis size ($d$) and a 3:2 axial ratio (to match that of the
EIS Patch D -- see below).  The source counts were normalised to provide an
average of 25 radio sources within a volume element defined by this field of
view ($d$ by $2d/3$ Mpc) and the length of the redshift interval $2.5 \lta z
\lta 3.5$ ($\sim 1000$\,Mpc). Then, for 100 different sight-lines through 
the simulation, the number of sources actually observed in this volume was
determined, and this was repeated for 150 different simulations of the
2000$^3$Mpc$^3$ volume. This whole process was carried out for a number of
different field sizes, $d$.

\begin{figure}
\centerline{
\psfig{file=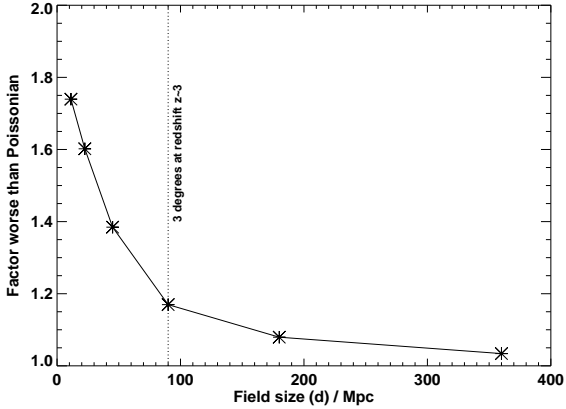,angle=90,width=8cm,clip=} 
}
\caption{\label{highzsize} The effect of radio source clustering on the high
redshift number counts. The plot shows the factor by which galaxy clustering
increases the width of the number count distribution relative to the
Poissonian expectation, as a function of field size. This is calculated by
constructing a source population with a spatial cross-correlation length of
12\,Mpc, normalised to have 25 sources over the field area and a unit step in
redshift ($2.5 < z < 3.5$). The field area is rectangular with a 3:2 axial
ratio, and the length of its longer axis ($d$) is shown as the abscissa. For
sky areas of 3$\times$2 degrees, the plot begins to level off and the effect
of large--scale structure is below the 20\% level.}
\end{figure}

As expected, the effect of large--scale structure is to broaden the observed
distribution of source counts, due to some sight--lines sampling clusters of
sources and others passing through voids. The effect of clustering was
quantified by measuring the width of the number count distribution, and
comparing it to the Poissonian expectation. The factor by which the
large--scale structure increases the distribution width (relative to
Poissonian) is plotted against the field--of--view, $d$, in
Figure~\ref{highzsize}. It is clear that for small fields clustering plays an
important role, but that when the field size gets up to about 100\,Mpc in size
the plot begins to level off and the effect of large--scale structure has
dropped below the 15--20\% level. This corresponds to an angular size of about
3 by 2 degrees (at $z \sim 3$) for the survey, which is the size of the ESO
Imaging Survey Patch D (see Section~\ref{sampdef}).

Finally, it is important to consider the number of radio sources required for
the survey. If there is no redshift cut-off then the Dunlop \& Peacock
\shortcite{dun90} models predict that over 20\% of radio sources in a
sample with 10\,mJy flux density limit should be at $z>2.5$, whilst if a $z>2$
decline is included then their pure luminosity evolution and luminosity
density evolution models predict respectively $\sim 9$ and $\sim 5$\% of
sources above that redshift. A sample of $\sim 150$ radio sources is
sufficient to distinguish between these possibilities at the $>3\sigma$ level,
whilst keeping the sample size to a manageable level. Conveniently, to obtain
this number of sources down to a $\sim 10$\,mJy flux density level also
requires a sky area of order six square degrees. The combination of the EIS
and NVSS surveys is therefore ideal.

\section{Sample Definition}
\label{sampdef}

\subsection{The ESO Imaging Survey (EIS)}

The wide--field ESO Imaging Survey (EIS-wide) comprises a relatively
wide--angle survey of four distinct patches of sky of up to 6 square degree
each, selected to have low optical extinction and to exclude especially bright
stars or nearby clusters of galaxies \cite{non99}. Of these four regions, the
most northerly, Patch-D, is a 3 by 2 degree field centred at RA: 09 51 36,
Dec -21 00 00 (J2000), and is thus sufficiently far north to be relatively
easily accessible for radio observations with the Very Large Array (VLA)
synthesis array. During the EIS-wide project, the entirety of Patch-D was
imaged in the $I$--band using the EMMI camera mounted on the 3.5-m New
Technology Telescope (NTT) at La Silla. This provides a pixel scale of 0.266
arcsec per pixel. Each region of sky was observed for 300 seconds in each of
two separate exposures.

A single--entry catalogue has been constructed for the EIS Patch D in
$I$--band \cite{non99}, in which objects detected in more than one overlapping
pointing are listed as a single entry, with the parameters determined from the
best seeing image; the seeing on the images ranges from 0.5 to 1.6 arcsec.
This catalogue contains over 560000 objects, reaching an 80\% completeness
limiting $I$ magnitude of typically $I \sim 23$ \cite{ben99}. Postage--stamp
images are also available. The EIS-wide survey of Patch D has been
complemented by a Wide Field Imager survey (using the ESO 2.2m telescope) of
the same field (see http://www.eso.org/science/eis/) in the $B$ and $V$ bands,
reaching limiting magnitudes of $V \sim 24.5$ and $B \sim 25$. Only catalogues
of each Wide Field Imager pointing are currently available, not a combined
single--entry catalogue nor postage stamp images. Astrometry on both of these
surveys is accurate to within 0.2 arcsec \cite{non99}.

The $B$, $V$, and $I$ filters used were filters designed especially for the
EIS project; they have effective wavelengths close to those of the
Johnson--Cousins $BVI$ filters, but are broader and with sharper cut-offs. The
transformation between the EIS magnitudes and the Johnson--Cousins magnitudes
is zero for stars with colours of zero, and is significantly below 0.1
magnitudes even for objects with the most extreme colours \cite{non99}.

The extracted magnitudes used in all wavebands were derived using SExtractor
(which was used to identify and measure properties of the galaxies in the EIS
catalogues; for details of SExtractor see Bertin \& Arnouts
1996)\nocite{ber96}, and are estimates of the total magnitude of the
object. Note, however, that these apertures have not been precisely matched
between the different catalogues, and so aperture difference will introduce
some scatter into the measured colours.

\subsection{The NRAO VLA Sky Survey (NVSS)}

The NRAO VLA Sky Survey (NVSS) is a radio survey carried out at a frequency of
1.4\,GHz using the Very Large Array in the D and DnC array configurations
\cite{con98}. This provides an angular resolution of about 45 arcseconds
full--width at half--maximum (FWHM). The survey covers the whole sky north of
$-40^{\circ}$ declination (J2000), to an root--mean--squared (rms) brightness
level of 0.45\,mJy\,beam$^{-1}$. Thus, the survey is essentially complete to
$\approx 3.5$\,mJy, to which level there are about 50 sources per square
degree.

The 45 arcsec beam size of the NVSS survey is significantly larger than the
median angular size of faint extragalactic sources at these flux density
levels ($\lta 10$\,arcsec; Condon \etal\ 1998)\nocite{con98}, meaning that
most sources will be unresolved. This has the benefit that for the majority of
detected objects an individual radio source is contained within a single NVSS
component, and the NVSS flux density measurement should have good photometric
accuracy with little or no flux density missed by resolving out
structure. However, the low angular resolution of this survey means that
follow--up observations are required in order to determine the structure of a
radio source, and to pin-point its position to allow identification of the
optical counterpart.

\subsection{Initial Sample Definition}

The EIS-NVSS radio source sample was initially defined in 1997, at which stage
only an early version of the NVSS catalogue had been produced, and the EIS
observations were still in progress. The {\it original} definition of this
sample was as follows:

\begin{itemize}
\item $09^{h} 44^{m} 25^{s} < {\rm RA} < 10^{h} 00^{m}$
\item $-22^{\circ} 6' < {\rm Dec} < -19^{\circ} 54'$
\item $S_{\rm 1.4\,GHz} \ge 7.8$\,mJy in original NVSS catalogue.
\end{itemize}

\noindent This corresponded to a sample of 199 radio sources, and these were
assigned a designation of `EISD' followed by a catalogue number which was `1'
for the brightest source and increased with decreasing NVSS flux density. The
sky area defined by these limits is slightly larger than the final $3 \times
2$ degree sky area of the EIS Patch D; it eventually turned out that 44 of
these sources lay outside of the region covered by the optical imaging, and so
are not considered in the final CENSORS sample discussed below. One source,
NVSS-J095759-2005 with $S_{\rm 1.4\,GHz} = 14.9 \pm 1.1$ was found to have
been accidentally excluded from the sample. Since this lies outside of the EIS
Patch D, and so is not within the CENSORS sample from which the radio
luminosity function work will be carried out, this exclusion will have no
significant consequences.

The sample properties differ somewhat from these initial selection criteria,
since the current public NVSS catalogue (version 2.17, issued July 2002)
differs substantially from the preliminary catalogue from which the observed
sample was selected. Significant deviations are found in both the positions
and the flux densities of the sources, suggesting that these have been
re-estimated. In general, the source flux densities have been amended
downwards, so that the new flux densities are on average nearly 10\% lower
than those originally selected, although with considerably scatter from source
to source. Importantly though, some of the sources that were just below the
flux density threshold in the original sample definition are now quoted to be
brighter than some sources that were included in the sample. The effects of
this are discussed in Section~\ref{finalsamp}, and the final CENSORS sample of
radio sources is defined there.

The properties of the 199 radio sources selected in the original EIS-NVSS
sample are provided in Table~\ref{obsdettab} of Appendix~\ref{radpropapp}.

\section{Radio Observations}
\label{obssect}

\subsection{VLA BnA array observations}

Radio observations of the 199 originally selected sources were taken at
1.4\,GHz using the VLA in BnA configuration on 15 and 19th June 1998 (see
Table~\ref{obsdettab} for details). The observations used the standard two IFs
at frequencies of 1385 and 1465\,MHz, with a bandwidth of 50\,MHz. The BnA
configuration is a hybrid between the B and A configurations with larger
spacings of the antennae along the northern arm, designed to produce
approximately circular beams for targets at low declinations. The angular
resolution provided by this setup is typically between 3 and 4 arcseconds.

Depending on their NVSS flux densities, the sources were observed for up to 8
minutes each. The exposure time for each source was set to provide a roughly
constant integrated signal--to--noise ratio of 60, for an unresolved
source. For those sources with total exposure times in excess of 5 minutes,
the exposure was split into two separate scans in order to improve the \uv\
coverage. The primary flux calibrator 3C286 (1331+305) was observed twice
during each of the runs, and was used to calibrate the flux density scale
assuming flux densities of 14.55 and 14.94\,Jy for that source at 1385 and
1465\,MHz respectively. These are the most recently determined VLA values
(1995.2), based upon the scale of Baars \etal\ \shortcite{baa77}. The nearby
secondary calibrator 0921$-$263 was observed approximately every 30 minutes to
provide accurate phase calibration.

The data were processed using standard techniques incorporated within the
\aips\ software provided by the National Radio Astronomy Observatory. After
first applying baseline corrections and discarding data from any antenna or
baseline showing excessive noise, the data were \clean ed using the \aips\
task \imagr. Maps were made over a 512 by 512 pixel field, with 0.6 arcsec
pixels; to reduce the noise from the sidelobes of sources outside this central
field and to obtain a good model of the field for subsequent
self--calibration, the positions of all other sources within the 30 arcmin VLA
primary beam with $S_{\rm 1.4GHz} > 7.5$\,mJy were extracted from the NVSS and
128 by 128 pixel fields centred on these positions were also included in the
\clean ing process. For 193 of the sources, sufficiently bright point sources
were observed within the fields that one or two cycles of phase
self--calibration were used to improve further the map quality.

\subsection{VLA CnB array observations}

As discussed below, a number of the more extended sources were largely
resolved out at the high angular resolution provided by the BnA configuration
observations. For these sources, radio data at lower resolution were
required. Observations of these sources (excepting EISD176 and EISD196 which
lay outside the region of EIS optical imaging, and were therefore excluded due
to telescope time constraints) were carried out in various observing runs
during 2002 (see Table~\ref{obsdettab} for details), with integration times of
typically about 5--10 minutes. Again, these observations were taken in two IFs
at frequencies of 1385 and 1465\,MHz, with a standard bandwidth of 50\,MHz.

These CnB array data were reduced in a similar way to the BnA array data,
except that in this case, due to the lower angular resolution, 1024 by 1024
pixel maps with 3 arcsec pixels were made to cover the entire primary
beam. After cleaning and self--calibration, the CnB array data were combined
with the BnA array data, a further cycle of self--calibration was carried out,
and final images of the sources were produced using both sets of data. The
resultant maps typically have angular resolutions of $\sim 7$ arcsec, but much
greater sensitivity to extended structures that the BnA array data alone.

\subsection{The radio maps}

Cleaned radio maps of all of the 199 radio sources have been produced. The rms
noise levels on each of these maps was calculated by taking the average of
five off--source regions, and is provided in Table~\ref{obsdettab}. For those
radio sources within the EIS field, contoured radio maps are shown in
Appendix~\ref{radpropapp}, in Figure~\ref{optovers}. These maps have contour
levels scaled in factors of 2 from a first contour level of three times the
rms noise on the final radio map. The full--width--half--maximum of the
Gaussian restoring beam is plotted in the lower left corner of each
map. Contour maps of the sources which lie outside of the EIS field are shown
in Figure~\ref{outovers}, with contour levels defined in the same way. No maps
are shown for those sources which are undetected in the radio waveband.

\section{Radio Properties of the Sources}
\label{radprops}

\subsection{Derived radio source parameters}

Source characteristics were derived from the radio maps; details of all of
these parameters are provided in Table~\ref{radsourprop}. The \aips\ task
\sad\ (Search And Destroy) was used, which fits Gaussian models to an image by
a least squares method, and estimates the errors \cite{con97b}. In this way,
peak flux densities ($S_{\rm peak}$), integrated component flux densities
($S_{\rm comp}$), and positions ($\alpha_{\rm comp}$, $\delta_{\rm comp}$) of
all of the source components were estimated. In addition, for resolved sources
the total flux density ($S_{\rm int}$) was calculated by integrating all of
the signal in a region surrounding the radio source; for unresolved source the
integrated flux density of the single source component was adopted instead.

The source morphologies were divided into five classes: single (S), double
(D), triple (T) and multiple (M) component sources, and extended diffuse
sources (E). For sources with more than one component, the position angle (PA)
and largest angular size ($D_{\rm rad}$) of the radio source were defined as
the position angle and the angular separation, respectively, of the two most
separated components. For single component sources, where the component is
resolved these properties were defined as the position angle and the major
axis length of the deconvolved elliptical Gaussian fit to the component. For
sources with a single unresolved component, the position angle was not
determined, and a 1$\sigma$ upper limit to the largest angular size was
derived from the upper limit to the major axis of the deconvolved elliptical
Gaussian fit.

\subsection{Notes on individual radio sources}

\noindent {\bf EISD7:} This source appears double with EISD44 in the NVSS
catalogue. The new radio map (Figure~\ref{eisd7fig}) indicates that EISD7 is
indeed a bright unresolved radio source, close to the position of its entry in
the NVSS catalogue. EISD44 is an extended radio source, stretching nearly 2
arcminutes on the sky, and therefore its NVSS catalogue position and flux
density are both inaccurate.

\begin{figure}
\caption{\label{eisd7fig} {\it See attached jpeg file}. The new radio map of
the EISD7 plus EISD44 combination. Radio contours are at
(-1,1,2,4,8,16,32,64,128,256,512,1024) $\times$ 66\,$\mu$Jy\,beam$^{-1}$.
EISD7 is an unresolved radio source, whilst the other radio components make
EISD44.}
\end{figure}

\noindent {\bf EISD15:} The weak second radio component, 25 arcsec
south-south-east of the central component, is probably unconnected with this
source.

\noindent {\bf EISD16:} This source overlaps with sources EISD56 and EISD114,
and a source below the 7.5\,mJy flux density limit in the NVSS catalogue
(Figure~\ref{eisd16fig}), and it is not entirely clear from the radio data
alone which radio components comprise which distinct sources. Addition of the
optical data (see Section~\ref{optids}) clears this up a bit, but not entirely
(Figure~\ref{eisd16fig}). It is clear that the radio components labelled `A'
on the figure correspond to an extended double radio source, associated with
EISD16. Component `F' is the weak radio point source seen in the NVSS map, and
has an associated optical counterpart. Component `D' also has an optical
counterpart, and causes the apparent extension of EISD56. The complication
arises with components `B', `C' and `E'. These may be simply interpreted as
three separate radio sources, of which `B' and `E' correspond to EISD114 and
EISD56 respectively, and `C' is a radio point source below the sample limit;
this component `C' also has an optical identification. However, the facts that
neither component `B' nor `E' has an optical counterpart, that `C' lies almost
midway between the two, and that component `B' apears to be slightly extended,
pointing towards `C', suggests an alternative explanation whereby these all
form one giant radio source with `C' as the core. The lack of extended radio
emission between the two in the NVSS map argues against this, and in favour of
the three radio source model; in the absence of evidence to the contrary, this
simple solution is adopted in the subsequent sample definition. However,
deeper radio observations will be required to elucidate exactly which
components constitute independent sources.

\begin{figure*}
\caption{\label{eisd16fig} {\it See attached jpeg file}. Upper left: the NVSS map of EISD16, EISD56 and
EISD114; radio contours are at ($-1$,1,2,4,8,16,32,64,128) $\times$
1.5\,mJy\,beam$^{-1}$.  Upper right: the new radio map produced by combining
the CnB and BnA array data for this field, overlaid upon the EIS $I$--band
image. Radio contours are at (-1,1,2,4,8,16,32,64,128,256,512,1024) $\times$
84\,$\mu$Jy\,beam$^{-1}$. Six different components of radio sources are
labelled; see text for discussion. Lower panels: postage-stamp enlargements of
the three components C,D and F (from left to right), demonstrating the relaity
of their optical counterparts.}
\end{figure*}

\noindent {\bf EISD20:} The NVSS entry for EISD20 corresponds to the bright
unresolved radio component to the south of the radio map in
Figure~\ref{optovers}. The further two components to the north appear as a
separate NVSS entry, whose flux density is too low to make it into the
EIS--NVSS sample. Although it is possible that these three comprise a single
extended radio source, the large flux density ratio between the two lobes in
this cases, coupled with the unresolved nature of the southern radio
component, make this unlikely. More likely, the unresolved component in the
south is indeed EISD20, as defined from the NVSS, whilst further north there's
a weak double hosted by the optical counterpart at 09 54 27.8, -21 56 27.

\noindent {\bf EISD24:} The additional two radio components to the south are 
most likely an unassociated double radio source.

\noindent {\bf EISD25:} In the NVSS catalogue, the sources EISD25 and EISD86
appear separated by less than 10 arcseconds. Figure~\ref{eisd25fig} shows
that these two NVSS sources result from a misfit to a single bright source
with a faint extension. The BnA array data clearly show that this is a single 
source, located at the peak of the NVSS flux. 

\begin{figure}
\caption{\label{eisd25fig} {\it See attached jpeg file}. A contour
representation of the NVSS map of EISD25 and EISD86, whose locations are
labelled by crosses, overlaid on a greyscale image of the new radio data. NVSS
radio contours are at ($-1$,1,2,4,8,16,32,64,128) $\times$
1.5\,mJy\,beam$^{-1}$. The two NVSS sources appear to arise from a misfit to
the NVSS data, due to the faint extension to the south--west. The new radio
data confirm this, with no emission detected at the location of EISD86.}
\end{figure}

\noindent {\bf EISD38:} The radio structure of this source is very unclear. A
narrow double is seen, together with a faint third component 25 arcsec further
north. The relationship between these components is unclear. The position
angle and largest size of the radio source were derived considering only the
southern components.

\noindent {\bf EISD44:} See comment on EISD7.

\noindent {\bf EISD56:} See comment on EISD16

\noindent {\bf EISD73:} EISD73 lies in an extended NVSS structure associated
with EISD151 and EISD103. Nearby on the sky, EISD77 is another extended NVSS
structure close to the source EISD112. These five radio sources are shown in
Figure~\ref{eisd73fig}. It is noteworthy that the two extended regions of
emission in the NVSS map are both oriented in the same direction, and that for
none of these three NVSS sources is any radio counterpart seen in the new CnB
or BnA images. Whilst it cannot be stated with 100\% confidence that these
sources do not represent extended emission, perhaps associated with an FR\,I
radio source, the failure to detect signal even in the CnB array data, and the
absence of a bright galaxy which might host such a radio source, both argue
against this. Further, the false source EISD118 (see below) is also found
nearby, and has similar orientation. These results strongly suggest that the
three sources associated with the extended NVSS emission regions (EISD73, 77
and 151) are not real, but instead are caused by correlated noise associated
with bad baselines or calibration errors in the NVSS data of that
pointing. These three sources are therefore removed from the final catalogue.

\begin{figure*}
\caption{\label{eisd73fig} {\it See attached jpeg file}. A contour
respresentation of the NVSS radio map of the EISD73, EISD77, EISD103, EISD112
and EISD151, with radio contours at ($-1$,1,2,4,8,16,32,64,128) $\times$
1.5\,mJy\,beam$^{-1}$. Overlaid on this is a greyscale image of the new BnA
radio data. The apparent extended emission associated with EISD73, EISD77 and
EISD151 is due to correlated noise on the NVSS map. Note that the CnB array
data (more sensitive to larger scale structure) also shows no evidence for
extended emission associated with these three sources.}
\end{figure*}

\noindent {\bf EISD77:} See comment on EISD73. This source is not real and 
is therefore removed from the final catalogue.

\noindent {\bf EISD86:} See comment on EISD25. This source is not real and 
is therefore removed from the final catalogue.

\noindent {\bf EISD90:} See comment on EISD113.

\noindent {\bf EISD98:} The weak second lobe to the north of this source is
very diffuse, but is believable because it provides consistency between the
observed position and that in the NVSS catalogue. The NVSS catalogue position
also coincides almost exactly with the position of a very bright optical
galaxy. 

\noindent {\bf EISD103:} See comment on EISD73

\noindent {\bf EISD112:} See comment on EISD73

\noindent {\bf EISD113:} The BnA radio map of this source has an odd
morphology.  This source was also observed off--axis in the CnB observations
of EISD7 and 44, although with significant primary beam attenuation. These
lower resolution observations confirm the two components extracted here (see
Figure~\ref{eisd113fig}). However, it is also noticeable that this radio
source lies close on the sky to EISD90, and it cannot be excluded that these
two form a single extended radio source. The lack of further extended emission
between the two and the compact nature of the EISD90 radio source, with a
potential optical identification, argue against this. The two sources are
therefore considered separately in this paper, although further radio
observations will be required to confirm this.

\begin{figure}
\caption{\label{eisd113fig} {\it See attached jpeg file}. A contour respresentation of the off--axis CnB
array radio map (corrected for primary beam attenuation) of EISD113 and
EISD90. Contour levels are at ($-1$,1,2,4,8,16,32,64,128) $\times$
200$\mu$Jy\,beam$^{-1}$. The $I$--band EIS image is overlaid in greyscale. } 
\end{figure}

\noindent {\bf EISD114:} See comment on EISD16

\noindent {\bf EISD118:} No radio source was detected. As shown in
Figure~\ref{eisd118fig}, this NVSS source appears to be just a residual close
to the bright sources EISD11 and EISD29, due to calibration and/or cleaning
errors. This source also lies close on the sky to the
EISD73--77--103--112--151 region, where the problems discussed above under
EISD73 are found, further confirming that this source is not real. It is
therefore removed from the final catalogue.

\begin{figure}
\caption{\label{eisd118fig} {\it See attached jpeg file}. A contour respresentation of the NVSS radio map of
EISD118, EISD11 and EISD29, with radio contours at ($-1$,1,2,4,8,16,32,64,128)
$\times$ 1.5\,mJy\,beam$^{-1}$, and a greyscale image of the new BnA radio
data overlaid. EISD118 is associated with an extended stripe of emission, with
no counterpart in the BnA array data, indicating that this source is not
real.}
\end{figure}

\noindent {\bf EISD124:} EISD124 and EISD137 are closely separated sources in
the NVSS catalogue, and the new radio maps (Figure~\ref{eisd124fig}) show
clearly that EISD137 is an unresolved source associated with a bright nearby
galaxy, whilst EISD124 stretches over 2.5 arcminutes top to bottom. The
northern component of EISD124 is unresolved from EISD137 in the NVSS image,
and thus the positions and locations of these two sources are poorly described
by their NVSS catalogue entries.

\begin{figure}
\caption{\label{eisd124fig} {\it See attached jpeg file}. The new radio map of the EISD124 plus EISD137
combination, overlaid upon the EIS $I$--band image. Radio contours are at
($-1$,1,2,4,8,16,32,64,128,256,512,1024) $\times$ 54\,$\mu$Jy\,beam$^{-1}$.
EISD137 is an unresolved radio source, whilst the other radio components make
up northern and southern lobes of the extended source EISD124.}
\end{figure}

\noindent {\bf EISD137:} See comment on EISD124

\noindent {\bf EISD151:} See comment on EISD73. This source is not real and is
therefore removed from the final catalogue.

\noindent {\bf EISD163:} This radio source was only weakly detected in the BnA
array observations, suggesting an extended radio source. Because the optical
images showed this to be clearly associated with a very nearby galaxy, more
detailed CnB radio data were not taken. For clarity, the NVSS contours are
used instead of the new BnA array data in the radio--optical overlay.

\noindent {\bf EISD176:} This radio source was undetected in the BnA array
observations but, because it lies outside the field of the optical imaging,
was not followed up in CnB array observations. It is likely to be a
significantly extended radio source.

\noindent {\bf EISD191:} As EISD163.

\noindent {\bf EISD196:} As EISD176.

\subsection{Radio source properties}

\begin{figure}
\centerline{
\psfig{file=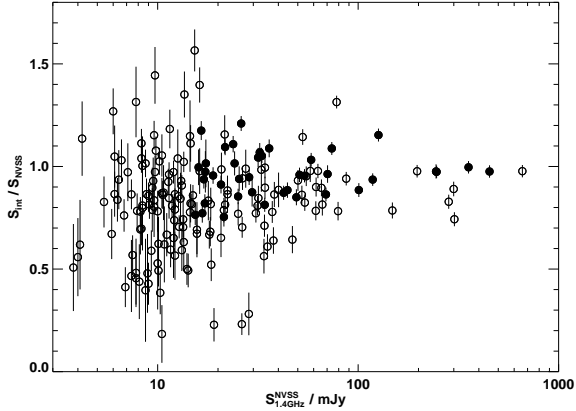,angle=90,width=8cm,clip=}
}
\caption{\label{fluxrat} The ratio of the 1.4\,GHz flux density determined for
the EIS sources in the new radio observations compared to the NVSS catalogue,
as a function of radio source flux density. Note that the new radio flux
densities have been increased by 1.5\% in this plot, to account for the small 
frequency difference (1425\,MHz compared to 1400\,MHz) of the current
observations compared to the NVSS, assuming a mean spectral index of 0.8. The
filled circles represent the single--component radio sources between EISD-1
and EISD-100, and the open circles are all fainter than this, or extended
sources.} 
\end{figure}

The radio flux densities of the EIS sources as determined in the new radio
observations are compared to the values in the NVSS catalogue in
Figure~\ref{fluxrat}. To make this comparison, the newly determined flux
densities were scaled up by $\sim 1.5$\% to account for the small difference
between the mean radio frequency of the current observations (1425\,MHz) and
those of the NVSS (1400\,MHz), assuming a typical radio spectral index of
$\alpha \approx 0.75$ (where $S_{\nu} \propto \nu^{-\alpha}$). It can be seen
that in addition to the scatter in the ratio of the two flux densities,
arising from the uncertainties in the two sets of measurements, there is also
a tendency for the new observations to determine a lower flux density than
suggested by the NVSS.

\begin{figure}
\centerline{
\psfig{file=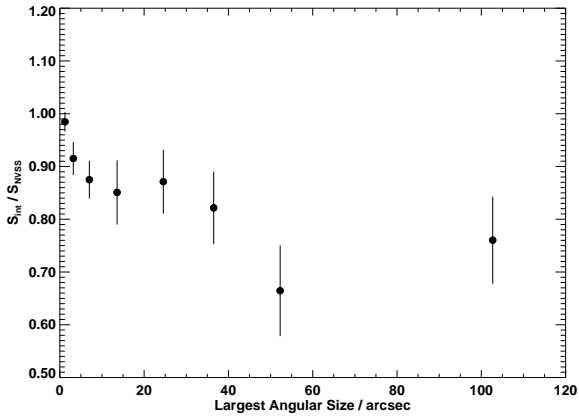,angle=90,width=8cm,clip=}
}
\caption{\label{lasflux} The median ratio of the 1.4\,GHz flux density
determined for the EIS sources in the new radio observations compared to the
NVSS catalogue, as a function of the angular size of the radio source, for 8
bins in angular size. The lower values for sources with larger angular extents
implies that some extended emission is being resolved out in the new radio
observations.}
\end{figure}

Sources which are in the brighter half of the sample, so that they are
detected at high signal--to--noise in both sets of observations, and which are
also unresolved, so that no flux should be lost in the current high resolution
observations, are indicated by filled circles in Figure~\ref{fluxrat}. The
NVSS flux densities of these are on average 2--3\% higher than those of the
new observations, which can be attributed to calibration errors between the
two different sets of observations. The flux densities scatter by of order
10\%, suggesting that the true uncertainties in the flux densities are
slightly higher than the formal errors in at least one of the two sets of
measurements. For fainter and extended sources (open circles in
Figure~\ref{fluxrat}) the average ratio is around 0.8 to 0.85, with larger
scatter; Figure~\ref{lasflux} demonstrates that the average flux ratio is
lower for sources with larger angular sizes, indicating that to at least some
extent this difference in flux densities is due to extended emission being
resolved out at the higher angular resolution of the new observations. The
NVSS flux densities, despite their larger uncertainties, probably give a more
accurate guide to the total flux densities of the extended sources.

The distribution of radio source morphologies across the five different
morphological classes defined above is indicated in Table~\ref{classnos}.
Fractionally over half of the sources are classified as single component
sources; this proportion rises to 60\% for the fainter sources (EISD101 to
EISD199), whilst the proportion of triple and multiple sources drops from 16\%
for the brightest 100 sources to 7\% for the fainter half of the sample. This
implies that at fainter radio flux density levels, a larger fraction of
unresolved radio sources are being picked out. 

\begin{figure}
\centerline{
\psfig{file=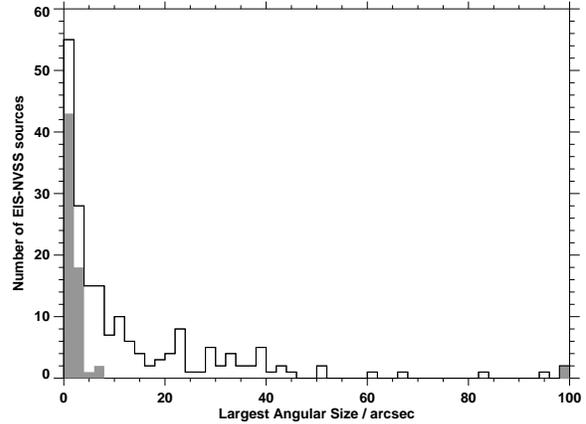,angle=90,width=8cm,clip=}
}
\caption{\label{linsize} The distribution of projected angular sizes of the
EIS-NVSS radio sources. The shaded regions represent upper limits to the
unresolved radio sources, and those sources whose angular sizes are larger
than 100\,arcsec.} 
\end{figure}

There is a possible concern that this results stems from second faint radio
components being missed for fainter radio sources, due to the relatively low
signal--to--noise of current the observations; it was shown above that
extended emission is often missed, and diffuse emission from an FR\,I source
will also be hard to detect. However, it is already well-known that fainter
radio sources tend to have smaller projected angular sizes (cf. Oort \etal
1987, Neeser \etal 1995)\nocite{oor87,nee95}, for example, the median
projected angular size of the revised 3CR sample ($\sim 100$ times brighter in
flux density that the current sample) is 43 arcsec, while that of the 6C
sample (a factor of $\sim 5$ fainter than 3CR) is only 14 arcsec
\cite{eal85a}. The distribution of projected angular sizes of the EIS-NVSS
radio sources is provided in Figure~\ref{linsize}: the median projected
angular size is found to be 6 arcsec, consistent with the above trend.  In
terms of multi-component nature, less than 10\% of the 3CR radio sources would
be unresolved at the resolution of the current observations \cite{lai83}, but
in the Hubble Deep Field observations of Richards \etal\ \shortcite{ric98a},
which reach about a factor of 100 deeper than the current sample, at least 25
of the 29 sources would be classified as single component sources. The
fraction of single component sources observed within the EIS sample, and the
trend with radio flux density, are both therefore consistent with these
results. This indicates that whilst there is a legitimate concern that a small
number of faint radio components may be missed, this should not dramatically
bias the properties of the sample as a whole.

\begin{table}
\caption{\label{classnos} Distribution of radio source morphologies}
\begin{tabular}{lccrrrr}
\hline
\multicolumn{2}{c}{Source structure} & ~~~~ & \multicolumn{4}{c}{Number of sources}  \\
  &  && All & \multicolumn{3}{c}{EISD1-100~~~EISD101-199} \\
\hline
Single   & (S) && 101 & ~~~~~~~~~42 & ~~~~~~~~~~~~59 \\
Double   & (D) && 66  & 39 & 27 \\
Triple   & (T) && 17  & 11 &  6 \\
Multiple & (M) && 6   &  5 &  1 \\
Extended & (E) && 2   &  0 &  2 \\
\hline
\multicolumn{2}{c}{Undetected/False} && 7 & 3 & 4 \\
\hline
\end{tabular}
\end{table}

\begin{figure}
\centerline{
\psfig{file=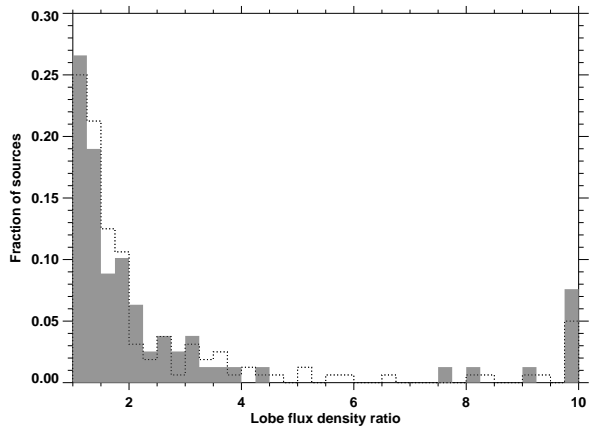,angle=90,width=8cm,clip=}
}
\caption{\label{loberat} The distribution of flux density ratios of the
brighter to the fainter lobe for those EIS-NVSS radio sources in which lobes
could be clearly defined (shaded histogram), compared to the equivalent values
for the 3CR sample (dotted histogram). The latter values are taken from McCarthy
\etal\ (1991).  The final bin contains all sources with flux density ratios in
excess of 10.}
\end{figure}
\nocite{mcc91}

An additional radio feature which is striking in some of the radio sources is
the dramatic difference in integrated flux density between the two different
lobes of the radio source. In cases such as EISD30 and EISD55, one lobe is
more than 20 times brighter than the other. In Figure~\ref{loberat} the flux
density ratios of the brighter to the fainter lobe are displayed for those
EIS-NVSS sources where the two lobes could be clearly defined; the median flux
density ratio is 1.7, with quite a long tail towards large asymmetries.
However, this feature is not unique to this sample, nor indeed to lower power
radio sources. McCarthy, van Breugel \& Kapahi \shortcite{mcc91} investigated
asymmetries in the 3CR sample, and their results (for radio galaxies and
quasars combined) are also shown on Figure~\ref{loberat}. Although at a
slightly higher radio frequency (5\,GHz), which may enhance beaming effects
and therefore increase asymmetries fractionally, it is clear that a comparable
range of asymmetries are seen in the 3CR sample as in the EIS-NVSS sample. The
median flux density ratio of the 3CR samples is slightly smaller, at 1.5, and
there are fewer sources with extreme asymmetries, but the differences between
the two samples are not statistically significant. It appears that there is 
little correlation between lobe flux density asymmetries and the flux density
of the radio source. A more detailed investigation of this property, and
investigation of further radio source asymmetries such as differences in the 
lobe lengths, and bending angles of the sources, must await the completion of
the host galaxy identifications of the sources.  

\section{Optical counterparts}
\label{optids}

\subsection{Radio--optical overlays}

3 by 3 arcminute $I$--band images around each of the radio sources were
retrieved from the EIS data archive. These co-added images use the conical
equal area projection (COE; e.g. Greisen \& Calabretta 2002)\nocite{gre02}
around the centre of the EIS Patch D at $\alpha = 147.00$, $\delta=-21.00$, in
order to produce equal area pixels with minimal distortion over the large 3 by
2 degree field. The COE projection is not supported by most data reduction
packages; it was therefore locally approximated by the standard TAN projection
by determining the exact pixel position on the image cut-outs appropriate for
the right ascension and declination of the radio source, and adjusting the
\fits\ headers so that the TAN projection gave the correct astrometry at this
position. However, the TAN projection approximation is only precise at the
position of the radio source: the offset between the TAN projection
approximation and the true position in these image cut-outs will increase with
distance from the radio source, reaching nearly a 1 arcsecond error at radial
distances of an arcminute. This is not important since in general only the
region at the very centre of the field around the radio source is of interest,
and in any case all optical positions were determined directly from the EIS
catalogue, where they are precisely calculated from the COE projection.

For all of the true radio sources which are within the EIS field, $I$--band
optical overlays are presented on the radio maps in Figure~\ref{optovers}.

\subsection{Identification likelihood ratios}

To investigate possible optical counterparts to the radio sources, a
likelihood analysis was carried out on the $I$--band catalogue. This was
carried out in two different ways. For single radio component sources, or
those in which a clear radio core is visible on the images, a simple
likelihood analysis based on the relative positions of the optical galaxies
and the radio position was carried out. This likelihood ratio technique
(e.g. Richter 1975, de Ruiter \etal\ 1977) can be used to statistically
investigate whether a proposed optical identification is the real counterpart
of a radio source, and is summarised in Appendix~\ref{appoptprop}. For
extended sources with no clear radio core, a modified likelihood analysis was
performed to account for the much greater uncertainty in the position of the
radio source. In this technique, the distribution of lobe arm--length ratios
and lobe bending angles for the 3CR sample (e.g. Best \etal\
1995)\nocite{bes95a} is used to provide a further prior on the expected host
galaxy position, and included in the likelihood analysis; this technique is
also described in Appendix~\ref{appoptprop}.

Details of the optical host candidates which had a likelihood ratio above the
cut-off for each of these two analysis are presented in Table~\ref{optprops}.
A further two classes of optical candidates were also added to this table. The
first of these are very bright host galaxies which were either excluded from
the EIS catalogue due to saturation, or were so large that the EIS catalogue
position is inaccurate leading to a low likelihood ratio. The second class of
added host candidates are optical galaxies which lie directly on top of an
unresolved ``lobe'' of extended radio sources, which would have a likelihood
ratio in excess of the cut-off if that radio component were deemed to be the
radio core. These were included due to the possibility that this unresolved
component is actually the AGN location, and the radio source is either
one--sided or a fainter second lobe has been missed in the new radio
observations. The selection method by which each optical host candidate is
included is noted in the table. In total, 102 of the 150 detected radio
sources within the EIS optical imaging field have likely optical
identifications. For the remainder, either the true optical host is detected
but the radio source is highly asymmetric and the identification falls below
the likelihood cut--off threshold, or the host galaxy is too faint to be
detected to this optical imaging depth.
 
For each of the 102 selected optical hosts, the $B$ and $V$ band catalogues
were searched for corresponding matches, and magnitudes of the host galaxies
in these bands, where detected, are presented in Table~\ref{optprops}. Also
presented in that table for each optical host is the `stellaricity' (S/G) of
the host galaxy in the $I$--band. This is a parameter measured by SExtractor
\cite{ber96} to help distinguish between stellar (or quasi--stellar) and
extended objects: unresolved objects are assigned a stellaricity value of 1.0,
and clearly resolved objects a value of 0.0, with the range of values in
between providing a likelihood estimate in uncertain cases.

\subsection{Notes on individual optical host galaxies}

\noindent {\bf EISD3:} In the EIS $I$--band catalogue the two central objects
erroneously appear as a single entry at 09 50 31.32, -21 02 44.4 with a
magnitude of 20.15. In the other two bands they appear as two separate
objects. The optical position was taken from the $V$--band data, and the 
$I$--band magnitude was re-determined.

\noindent {\bf EISD6:} Two potential optical counterparts are found to 
this extended radio source. These have similar likelihood ratios, with the
brighter galaxy to the north marginally favoured.

\noindent {\bf EISD16:} Although there is a weak radio feature towards the
centre of this radio source, it is not clear if this is a core. A likelihood
analysis treating the source as an extended source picks out the unresolved
object associated with a radio structure to the north as the most likely
optical counterpart.

\noindent {\bf EISD38:} The $I \sim 19$ galaxy lying exactly on the
south--eastern radio component is selected as a potential host galaxy.

\noindent {\bf EISD41:} There are two optical objects identified between the
radio lobes, but neither achieves a likelihood ratio in excess of the
cut-off. 

\noindent {\bf EISD47:} This candidate should be considered very tentative,
owing to the extremely faint $I$--band magnitude.

\noindent {\bf EISD60:} The faint optical object associated with the western
radio component is kept as a potential id.

\noindent {\bf EISD87:} There are many optical galaxies close to the radio
position, but none achieves a likelihood ratio above the cut-off value.

\noindent {\bf EISD91:} The two optical objects near the centre of the radio
source are blended into a single object in the $I$--band catalogue. In the other
two bands they appear as two separate objects. The optical position was taken
from the $V$--band data, and the $I$--band magnitude was re-determined.

\noindent {\bf EISD102:} The luminous galaxy coincident with the southern
radio component is retained as a likely optical counterpart.

\noindent {\bf EISD110:} Strangely, this bright optical galaxy is not found
within the EIS $I$--band catalogue, although it is in the $V$ and $B$ band
catalogues. Its $I$--band magnitude was measured directly from the image.

\noindent {\bf EISD120:} The faint extended emission seen in the $I$--band
overlay is not catalogued within the EIS catalogue, but appears convincingly
on the image and so its magnitude and position have been determined.

\noindent {\bf EISD123:} This bright saturated galaxy is excluded from the
EIS catalogue, but is clearly the optical counterpart to the radio source. The
optical image shows it to be part of a merging system, with numerous bright
tidal features, including a long tidal tail stretching to the north. 

\noindent {\bf EISD148:} The $I \sim 19$ object coincident with the eastern
radio component is considered a likely optical id.

\noindent {\bf EISD155:} There is a bright galaxy overlying the northern radio
component, which is identified as a likely counterpart.

\noindent {\bf EISD162:} The likely radio source counterpart is a fainter
object close to a brighter star. Although clearly distinct, these two objects
are not separated in any of the three catalogues. In the $I$--band the position
and an estimate of the magnitude of the source were determined from the image.

\noindent {\bf EISD163:} This very bright saturated galaxy is not included in
the optical catalogue. Also known as ESO 566-G 014, it has a redshift of
0.01559 in the NASA/IPAC Extragalactic Database (NED). Its magnitudes quoted
there are $B=14.9 \pm 0.1$ , $R=14.0 \pm 0.1$ and $I=13.7 \pm 0.1$.

\noindent {\bf EISD171:} Two plausible optical counterparts are picked out by
the likelihood analysis, although in fact there are several $I$--band objects 
close to the radio lobes, the majority of which (including the two selected
candidates) appear to be unresolved. It is not clear whether either of these
is really a good candidate.

\noindent {\bf EISD178:} The galaxy close to the south--western radio lobe is
retained as a plausible optical host.

\noindent {\bf EISD181:} This candidate should be considered very tentative,
owing to the extremely faint $I$--band magnitude.

\noindent {\bf EISD191:} This very bright nearby spiral galaxy has a redshift
of $z=0.02935$ in NED, but in the EIS imaging is saturated and excluded from
the catalogue. The galaxy is also known as ESO 566-G 018, and has colours from
NED of $B=14.1 \pm 0.1$, $R=13.1 \pm 0.1$ and $I = 12.7 \pm 0.1$.

\subsection{The optical hosts}

\begin{figure}
\centerline{
\psfig{file=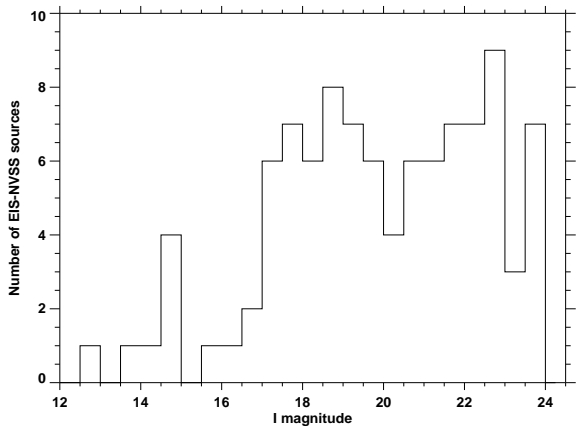,angle=90,width=8cm,clip=}
}
\caption{\label{ihist} The distribution of $I$--band magnitudes of the objects
selected as likely optical counterparts of the EIS-NVSS radio sources. The 48
radio sources without a sufficiently likely optical counterpart are not
represented on this plot. The majority of these are likely to be undetected in
the EIS imaging, and hence have $I \gta 23.5$, but some may simply be
asymmetric radio sources for which the true host is optically detected but is
not selected as having a sufficiently high likelihood ratio.}
\end{figure}

The distribution of $I$--band magnitudes of the galaxies selected as likely
optical counterparts to the radio sources is shown in Figure~\ref{ihist}. For
48 radio sources, no optical counterpart with a sufficiently high likelihood
ratio was found, and so these are not represented on this plot. It is likely
that the majority of these lie fainter than the $I \sim 23.5$ magnitude limit
of the EIS observations, but some may be brighter host galaxies which are
detected in EIS but do not have sufficiently high likelihood ratios, for
example due to highly asymmetric radio sources.

The $I$--band magnitudes can be used to provide a first estimate of the
redshift distribution of the radio sources\footnote{This method is preferred
to using the 3-colour information to obtain photometric redshifts, because on
comparison with our existing spectroscopic data (Brookes \etal\ in
preparation) it is found to be more reliable, even when the template models in
the photometric redshift estimation are restricted to only passive
ellipticals. This is undoubtedly due to the small number of available colours,
the fact that apertures are not explicitly matched, and the generally large
uncertainties on the B-band magnitudes.}. This is because powerful radio
sources are invariably hosted by giant elliptical galaxies with a narrow
spread of absolute magnitudes, and hence their magnitudes and redshifts are
tightly correlated. These correlations show least scatter at near--infrared
wavelengths (e.g. the $K-z$ relation; Lilly \& Longair 1984)\shortcite{lil84a}
where the emission is dominated by that of old evolved stars, even at
redshifts $z \sim 1-2$, and so any on--going low--level star formation or
emission associated with the active nucleus have lesser effect.

Magnitude--redshift correlations are also found at optical wavelengths
(e.g. Eales 1985c)\nocite{eal85b} although these have more scatter,
particularly for powerful radio galaxies at high redshifts. This is because
powerful radio galaxies with redshift $z \gta 0.6$ display considerable excess
blue emission aligned along their radio axes, due to AGN--related activity or
recent star formation (the alignment effect; McCarthy \etal\ 1987, Chambers
\etal\ 1987).\nocite{mcc87,cha87} The use of an $R-z$ or $I-z$ relation for
3CR sources beyond that redshift is therefore not appropriate for the
EIS--NVSS sources, which are expected to be much more passive because the
strength of the alignment effect is strongly correlated with radio power
\cite{ins03a}. Gigahertz--peaked spectrum (GPS) sources, however, show no
strong alignment effect, and so their optical magnitude versus redshift
relation may provide a good approximation to that of the EIS--NVSS
sources. Snellen \etal\ \shortcite{sne96a} find that the $r-z$ relation of GPS
radio galaxies can be roughly parameterised as $r = 22.7 + 7.4{\rm log}z$.

Magnitude--redshift relations only hold for the radio galaxies, and so the
quasars must be removed from the sample. Table~\ref{optprops} gives the $BVI$
colours of the optical candidates, and an estimate of the star--galaxy
classification of these in the $I$--band, using the SExtractor stellaricity
estimator provided in the EIS catalogue. This estimator gives a value ranging
between 0 (for galaxies) and 1 (for stars). Of the 102 optical counterparts,
the 12 with a stellaricity S/G$>$ 0.9 were classified as likely quasars, as
were the further two objects with stellaricities S/G$>$ 0.6 and colours $B-I <
1.0$, since distant galaxies are unlikely to have colours this blue (the other
objects with 0.6 $<$S/G$<$ 0.9 all had colours $B-I > 1.8$, so the exact
choice of colour cut-off is not critical). Note that this fraction of quasars
($\sim 14$\%) is much lower than the $\sim 30$\% found in the brightest radio
source samples (e.g. Best \etal\ 1999)\nocite{bes99e}, although some fainter
quasars may have been misclassified as galaxies if their signal--to--noise is
too low to allow good differentiation.

\begin{figure}
\centerline{
\psfig{file=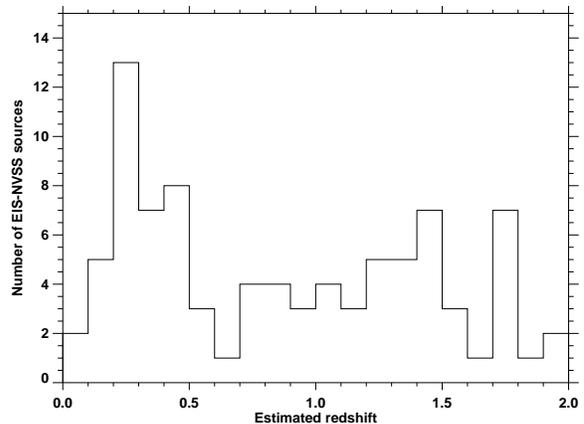,angle=90,width=8cm,clip=}
}
\caption{\label{zhist} The estimated redshift distribution of the 89 radio
sources with optical identifications that are not associated with suspected
quasars. This redshift distribution is based solely upon the $I$--band
magnitudes (see text for details). As in Figure~\ref{ihist}, the 48 radio
sources without a sufficiently likely optical counterpart are not represented
on this plot. The majority of these will be optically undetected to $I \sim
23.5$, and hence have estimated redshifts $z_{\rm est} \gta 1.75$.}
\end{figure}

The $I$--magnitudes of the radio galaxy hosts were converted to
$r$--magnitudes using the K and evolutionary corrections of Poggianti
\shortcite{pog97}, for an elliptical galaxy which formed at high redshift with
an exponentially decreasing star formation rate of e--folding time of
1\,Gyr. These were then used to provide a redshift estimate using the
parameterised $r-z$ relation above. Such conversions are undoubtedly uncertain
at the $\pm 0.2$magnitude level, but are accurate enough to allow rough
redshift estimation and hence a first look at the redshift distribution of the
EIS--NVSS sources. The estimated redshift distribution is shown in
Figure~\ref{zhist}. Again, the 48 sources with no acceptable optical
identification are excluded from this plot (as are the quasars for which
redshift estimation was not possible). A large proportion of these 48
unidentified sources are likely to have $I \gta 23.5$, and hence lie at
redshifts $z \gta 1.5-2$. Over a third of the EIS-NVSS sources are therefore
estimated to lie at high redshifts, $z \gta 1.5$; spectroscopic measurements
are clearly required, but this sample should indeed prove ideal for
investigation of the reality of a high redshift decline of the space density
of powerful radio sources.

\section{The CENSORS sample}
\label{finalsamp}

\subsection{Sample definition}

As discussed in Section~\ref{sampdef}, the flux densities of the sources in
the NVSS catalogue have been re-estimated since the original sample was
defined. In addition, 44 of the EIS--NVSS radio sources lie outside of the EIS
Patch D, five of the NVSS sources within the optical imaging region have been
shown to be false, and some extended radio sources extend over more than one
NVSS component. These facts require a re--definition of the final sample of
radio sources for further study, which hereafter is referred to as the {\it
CENSORS} sample ({\it A Combined EIS--NVSS Survey Of Radio
Sources})\footnote{Note that this CENSORS numbering system was not used from
the start of this paper, because the EIS--NVSS numbering system is that which
corresponds to the VLA radio observations in the archive, and also to the
majority of the infrared and spectroscopic follow--up (Brookes \etal\ in
preparation), and hence will be required for archive researchers.}

150 of the observed radio sources satisfy the criteria of the CENSORS sample,
namely being confirmed radio sources within the EIS imaging area, and the new
sample is defined in Table~\ref{finalsamptab}, in Appendix~\ref{appfinalsamp}.
These sources are ordered in order of decreasing flux density in the latest
version (v2.17) of the NVSS catalogue, with adjustments to these values being
made for five sources, marked with asterisks, due to their overlap with other
NVSS sources: for EISD7 and EISD137, the flux densities of the point sources
are accurately determined from the new observations, converted to a 1.400\,GHz
flux density assuming a spectral index of 0.8, and the NVSS flux density not
accounted for in these point sources is assigned to EISD44 and EISD124
respectively. For EISD56 the flux density is calculated excluding the
contribution from the nearby point sources. The median 1.4\,GHz flux density
of the CENSORS sample is $\approx 15$\,mJy.

\subsection{Sample Completeness}

The revision in the NVSS flux densities means that there is now a tail of
sources with low flux densities. Considering the most recent NVSS catalogue,
the CENSORS sample includes all of the radio sources brighter than $S_{\rm
1.4\,GHz} = 7.2$\,mJy, together with additional fainter sources. Further, only
three sources brighter than 6.5\,mJy were not targetted in the original EISD
radio observations, these being NVSS-J094651-2125 at $S_{\rm 1.4\,GHz} =
7.2$\,mJy, NVSS-J095233-2129 at 6.8\,mJy and NVSS-J095240-2123 at 6.7\,mJy.
All three of these sources lie well within the primary beam of the radio
observations of a different EIS--NVSS source, and so radio data are available
in each case. Their radio--optical overlays are shown in
Figure~\ref{extrasfig}, and their properties are included in
Table~\ref{finalsamptab} as CENSORS-X1 to X3. Including these three sources in
the CENSORS sample makes it complete for all NVSS sources brighter than
$S_{\rm 1.4\,GHz} = 6.5$\,mJy, with 9 additional sources below this limit.

Another recent development is that Blake \& Wall \shortcite{bla02b} showed
that about 7\% of all radio sources are resolved into multiple components by
NVSS. This multiple--component effect is seen within the NVSS sources selected
in the current sample, as discussed above. However, it is clearly also
important to consider whether radio sources which should be above the flux
density limit have been excluded from the sample due to being split into
different NVSS components, each of which fell below the flux density
limit. This was tested in two ways. Firstly the entire NVSS map of the EIS
field was eyeballed to search for potential extended sources with more than
two NVSS components, or of physical extent in excess of 3 arcminutes. No
promising candidates were found (it should be noted that Lara \etal\
\shortcite{lar01} estimated that only 1 in 10000 NVSS radio sources are larger
than 3 arcmins in size, and so in a small field like this, none would be
expected). Secondly, using the NVSS catalogue of this region, a search was
made for all NVSS source pairs separated by less than 3 arcminutes, where the
two NVSS sources were both fainter than 6.5\,mJy but the combined flux would
put them in to the sample. From the results of Blake \& Wall
\shortcite{bla02b} about 15\% of all pairs with 3 arcmin separations are
expected to be true double sources, rising to nearly 50\% at 2 arcmins, and
the vast majority of smaller separation doubles.

Only two NVSS pairs not already included in the sample were found with
separations below 2 arcmins. One of these is clearly two distinct radio
sources, since one of the sources has a bright optical counterpart.  For the
other, neither NVSS source has a clear optical counterpart, but one of the
NVSS sources appears off-centre in one of the BnA radio maps, and appears
unresolved. Also, there is no evidence in the NVSS map for any extension
between the sources. Therefore it appears that both of these two pairs are
chance alignments. A further 6 NVSS source pairs are found with separations
between 2 and 3 arcmins. Three of these can be excluded as doubles on the
basis of the radio and optical data; a further two are ambiguous, with no
bright optical identifications for either NVSS source, but insufficient
positional accuracy to rule out faint counterparts. However, these have no
apparent extensions in the NVSS, no obvious bright candidate host galaxy
between the two radio lobes (the majority of large angular separation radio
sources have relatively low redshifts), and flux density ratios $\gta 2$
between the two NVSS sources, and so are unlikely to be double radio
sources. These source pairs are therefore not considered further. The final
NVSS source pair, however, NVSS-J095218-2038 and NVSS-J095223-2041 with flux
densities of 3.6 and 3.8mJy and a separation of 2.8 arcmins, provide a
promising candidate (cf Figure~\ref{extrasfig}). These sources show extensions
in the NVSS indicative of a possible double, have a flux density ratio close
to unity, have no obvious optical counterparts individually, but have an $I
\sim 22$ galaxy situated roughly midway between them. This pair is therefore
provisionally added to the final sample (called CENSORS-D1) as a possible
large double.

\section{Conclusions}
\label{concs}

A new sample of $\sim 150$ radio sources has been defined by combining the
NVSS radio survey with the EIS optical observations over a 3 by 2 degree patch
of sky. The median radio flux density of these sources is $S_{\rm 1.4\,GHz}
\sim 15$\,mJy, corresponding to a factor of a few lower flux densities than
existing samples of radio sources with complete redshift information. These
sources are selected at an ideal flux density limit to carry out a census of
the high redshift space density of radio sources, and specifically to
investigate the reality of a high redshift cut--off in the radio luminosity
function. It is shown that the field studied is large enough that the effects
of radio source clustering will not dominate the results.

High resolution VLA radio observations have cleared up ambiguities in the
radio selected sample and provided accurate positions and morphologies for
these sources. The median radio size of the sources is 6 arcsec, and
approximately half of the sources appear as single radio component sources;
these results are in line with the trend for lower flux density radio sources
to be typically less extended. The lobe flux ratio asymmetries of the sources
are comparable to those of the brightest radio sources.

Comparison with the EIS optical imaging has allowed optical counterparts to be
identified for over two--thirds of the radio sources. Estimated redshifts for
these sources, based upon their optical magnitudes, suggest that over a third
of the galaxies lie at redshifts $z \gta 1.5$ confirming the interest in
spectroscopic follow--up of this sample. Infrared imaging to identify the
remainder of the sample and the results of spectroscopic follow--up
observations will be presented in subsequent papers.

\section*{Acknowledgements} 

PNB would like to thank the Royal Society for generous financial support
through its University Research Fellowship scheme.  MHB is grateful for the
support of a PPARC research studentship. The authors wish to thank the
referee, Mark Lacy, for some useful comments. The National Radio Astronomy
Observatory is a facility of the National Science Foundation operated under
cooperative agreement by Associated Universities, Inc. The observations of the
EIS--WIDE survey were carried out using the ESO New Technology Telescope (NTT)
at the La Silla observatory under Program-ID Nos. 59.A-9005(A) and
61.A-9005(A); those of WFI Pilot Survey were carried out using the MPG/ESO
2.2m Telescope at the La Silla observatory under Program-ID No. 163.O-0741.
This research has made use of the NASA/IPAC Extragalactic Database (NED) which
is operated by the Jet Propulsion Laboratory, California Institute of
Technology, under contract with NASA.

\bibliography{pnb} 
\bibliographystyle{mn}
\appendix

\section{Properties of the Radio Sources}
\label{radpropapp}

In this appendix, properties of the radio sources are provided.
Table~\ref{obsdettab} provides details of the original EIS--NVSS sample of
radio sources, and the observations of these. Radio maps of those sources
which are located within the region of optical imaging of the EIS survey are
shown in Figure~\ref{optovers}, overlaid upon greyscale images taken from the
$I$--band EIS data. Radio maps of those which lie outside of the EIS Patch D
are shown in Figure~\ref{outovers}. Table~\ref{radsourprop} provides the
properties of the radio source components, as derived from these new radio
maps.

\begin{table*}
\caption{\label{obsdettab} Details of the EIS-NVSS sample of radio sources.
The flux densities quoted are at 1.4\,GHz as extracted from the most recent
version of the NVSS catalogue (version 2.17, issued July 2002). The rms noise
is that on the final image produced from the new VLA observations (after
combination of array configurations where appropriate) presented in this
paper.}
\begin{tabular}{lcccrccr}
\hline
\multicolumn{1}{c}{Source} &
IAU Name &
\multicolumn{2}{c}{RA~~~(NVSS)~~~Dec}      &
\multicolumn{1}{c}{$S_{\rm 1.4\,GHz}^{\rm NVSS}$} &
\multicolumn{2}{c}{Observation Date} &
\multicolumn{1}{c}{rms} \\
&
&
\multicolumn{2}{c}{(J2000)} &
\multicolumn{1}{c}{[mJy]} &
BnA array &
CnB array &
\multicolumn{1}{c}{[$\mu$Jy]} \\
\hline
EISD1   & NVSS-J095129-2050 & 09 51 29.17 & -20 50 28.6 & 659.5 $\pm$ 19.8 & 15/06/1998 &     ---    &  65 \\
EISD2   & NVSS-J094650-2020 & 09 46 50.28 & -20 20 44.1 & 452.3 $\pm$ 13.6 & 15/06/1998 &     ---    &  58 \\
EISD3   & NVSS-J095031-2102 & 09 50 31.40 & -21 02 43.0 & 355.3 $\pm$ 10.7 & 15/06/1998 &     ---    &  49 \\
EISD4   & NVSS-J094905-1957 & 09 49 05.67 & -19 57 10.6 & 302.1 $\pm$ 10.0 & 15/06/1998 &     ---    &  35 \\
EISD5   & NVSS-J095953-2148 & 09 59 53.89 & -21 48 47.7 & 299.6 $\pm$ 10.5 & 15/06/1998 &     ---    &  47 \\
EISD6   & NVSS-J094953-2156 & 09 49 53.57 & -21 56 17.4 & 283.0 $\pm$  9.5 & 15/06/1998 &     ---    &  33 \\
EISD7   & NVSS-J095143-2123 & 09 51 43.69 & -21 23 56.4 & 246.2 $\pm$  7.4 & 15/06/1998 & 30/04/2001 &  32 \\
EISD8   & NVSS-J095344-2135 & 09 53 44.01 & -21 35 51.7 & 244.7 $\pm$  8.2 & 15/06/1998 &     ---    &  34 \\
EISD9   & NVSS-J095812-2144 & 09 58 12.96 & -21 44 41.0 & 196.9 $\pm$  5.9 & 15/06/1998 &     ---    &  48 \\
EISD10  & NVSS-J094557-2116 & 09 45 57.05 & -21 16 48.2 & 148.2 $\pm$  5.1 & 15/06/1998 &     ---    &  42 \\
EISD11  & NVSS-J095730-2130 & 09 57 30.00 & -21 30 58.2 & 126.3 $\pm$  3.8 & 15/06/1998 &     ---    &  44 \\
EISD12  & NVSS-J094935-2156 & 09 49 35.38 & -21 56 23.2 & 118.2 $\pm$  3.6 & 15/06/1998 &     ---    &  40 \\
EISD13  & NVSS-J094427-2116 & 09 44 27.26 & -21 16 10.8 & 100.8 $\pm$  3.1 & 15/06/1998 &     ---    &  48 \\
EISD14  & NVSS-J095951-2053 & 09 59 51.08 & -20 53 18.9 &  87.2 $\pm$  2.7 & 15/06/1998 &     ---    &  38 \\
EISD15  & NVSS-J095329-2002 & 09 53 29.54 & -20 02 13.0 &  78.1 $\pm$  2.4 & 15/06/1998 &     ---    &  40 \\
EISD16  & NVSS-J094727-2126 & 09 47 27.03 & -21 26 21.0 &  79.4 $\pm$  2.9 & 15/06/1998 & 30/04/2001 &  28 \\
EISD17  & NVSS-J094433-2105 & 09 44 33.17 & -21 05 07.2 &  73.9 $\pm$  2.3 & 15/06/1998 &     ---    &  45 \\
EISD18  & NVSS-J094641-2029 & 09 46 41.16 & -20 29 26.0 &  70.4 $\pm$  2.6 & 15/06/1998 &     ---    &  47 \\
EISD19  & NVSS-J094508-2038 & 09 45 08.44 & -20 38 07.9 &  69.1 $\pm$  2.1 & 15/06/1998 &     ---    &  42 \\
EISD20  & NVSS-J095428-2156 & 09 54 28.91 & -21 56 53.6 &  66.3 $\pm$  2.7 & 15/06/1998 &     ---    &  29 \\
EISD21  & NVSS-J095447-2059 & 09 54 47.70 & -20 59 43.4 &  65.6 $\pm$  2.4 & 15/06/1998 &     ---    &  43 \\
EISD22  & NVSS-J094651-2053 & 09 46 51.17 & -20 53 17.5 &  63.0 $\pm$  1.9 & 15/06/1998 &     ---    &  35 \\
EISD23  & NVSS-J095751-2133 & 09 57 51.31 & -21 33 21.5 &  61.7 $\pm$  2.3 & 15/06/1998 &     ---    &  60 \\
EISD24  & NVSS-J095242-1958 & 09 52 42.99 & -19 58 20.0 &  61.5 $\pm$  2.3 & 15/06/1998 &     ---    &  32 \\
EISD25  & NVSS-J095513-2123 & 09 55 13.61 & -21 23 03.2 &  58.3 $\pm$  1.8 & 15/06/1998 &     ---    &  32 \\
EISD26  & NVSS-J095904-2008 & 09 59 04.87 & -20 08 05.0 &  57.7 $\pm$  2.2 & 15/06/1998 &     ---    &  37 \\
EISD27  & NVSS-J095330-2135 & 09 53 30.49 & -21 35 58.9 &  55.1 $\pm$  2.1 & 15/06/1998 &     ---    &  46 \\
EISD28  & NVSS-J094758-2121 & 09 47 58.99 & -21 21 50.6 &  54.0 $\pm$  1.7 & 15/06/1998 &     ---    &  41 \\
EISD29  & NVSS-J095730-2132 & 09 57 30.82 & -21 32 37.7 &  52.9 $\pm$  1.7 & 15/06/1998 &     ---    &  32 \\
EISD30  & NVSS-J094604-2115 & 09 46 04.78 & -21 15 08.8 &  54.2 $\pm$  2.1 & 15/06/1998 &     ---    &  36 \\
EISD31  & NVSS-J095629-2001 & 09 56 29.93 & -20 01 30.5 &  52.4 $\pm$  2.0 & 15/06/1998 &     ---    &  32 \\
EISD32  & NVSS-J095438-2104 & 09 54 38.38 & -21 04 25.3 &  51.0 $\pm$  1.6 & 15/06/1998 &     ---    &  46 \\
EISD33  & NVSS-J095433-2205 & 09 54 33.72 & -22 05 22.2 &  50.1 $\pm$  1.9 & 15/06/1998 &     ---    &  40 \\
EISD34  & NVSS-J094804-2147 & 09 48 04.06 & -21 47 36.5 &  49.2 $\pm$  1.9 & 15/06/1998 &     ---    &  33 \\
EISD35  & NVSS-J095902-2039 & 09 59 02.43 & -20 39 45.7 &  47.0 $\pm$  2.1 & 15/06/1998 &     ---    &  42 \\
EISD36  & NVSS-J095217-2008 & 09 52 17.76 & -20 08 35.3 &  44.4 $\pm$  1.4 & 15/06/1998 &     ---    &  29 \\
EISD37  & NVSS-J095825-2044 & 09 58 25.97 & -20 44 51.3 &  42.5 $\pm$  1.4 & 15/06/1998 &     ---    &  31 \\
EISD38  & NVSS-J094631-2026 & 09 46 31.42 & -20 26 10.2 &  40.1 $\pm$  1.9 & 15/06/1998 &     ---    &  46 \\
EISD39  & NVSS-J094815-2140 & 09 48 15.71 & -21 40 05.1 &  38.2 $\pm$  1.6 & 15/06/1998 &     ---    &  36 \\
EISD40  & NVSS-J094556-2028 & 09 45 56.53 & -20 28 31.3 &  37.8 $\pm$  2.0 & 15/06/1998 & 30/04/2001 &  35 \\
EISD41  & NVSS-J094519-2142 & 09 45 19.62 & -21 42 39.5 &  37.3 $\pm$  1.5 & 15/06/1998 &     ---    &  45 \\
EISD42  & NVSS-J095827-2105 & 09 58 27.40 & -21 05 26.5 &  36.0 $\pm$  1.2 & 15/06/1998 &     ---    &  35 \\
EISD43  & NVSS-J095141-2011 & 09 51 41.06 & -20 11 17.9 &  35.3 $\pm$  1.5 & 15/06/1998 &     ---    &  25 \\
EISD44  & NVSS-J095150-2125 & 09 51 50.39 & -21 25 14.4 &  33.9 $\pm$  2.3 & 15/06/1998 & 30/04/2001 &  22 \\
EISD45  & NVSS-J095304-2044 & 09 53 04.75 & -20 44 09.6 &  34.3 $\pm$  1.1 & 15/06/1998 &     ---    &  33 \\
EISD46  & NVSS-J094455-2017 & 09 44 55.53 & -20 17 14.6 &  34.3 $\pm$  1.4 & 15/06/1998 &     ---    &  45 \\
EISD47  & NVSS-J094753-2147 & 09 47 53.64 & -21 47 19.2 &  34.2 $\pm$  1.1 & 15/06/1998 &     ---    &  34 \\
EISD48  & NVSS-J095452-2119 & 09 54 52.43 & -21 19 28.8 &  34.1 $\pm$  1.4 & 15/06/1998 &     ---    &  34 \\
EISD49  & NVSS-J095925-2006 & 09 59 25.82 & -20 06 40.3 &  32.9 $\pm$  2.0 & 15/06/1998 &     ---    &  39 \\
EISD50  & NVSS-J095724-2203 & 09 57 24.19 & -22 03 52.6 &  33.1 $\pm$  1.1 & 15/06/1998 &     ---    &  44 \\
EISD51  & NVSS-J094933-2127 & 09 49 33.26 & -21 27 07.7 &  32.3 $\pm$  1.1 & 15/06/1998 &     ---    &  29 \\
EISD52  & NVSS-J094919-2151 & 09 49 19.33 & -21 51 35.4 &  31.8 $\pm$  1.4 & 15/06/1998 &     ---    &  35 \\
EISD53  & NVSS-J095116-2056 & 09 51 16.85 & -20 56 35.2 &  31.7 $\pm$  1.1 & 15/06/1998 &     ---    &  50 \\
EISD54  & NVSS-J094836-2106 & 09 48 36.08 & -21 06 23.0 &  31.5 $\pm$  1.1 & 15/06/1998 &     ---    &  32 \\
EISD55  & NVSS-J095058-2114 & 09 50 58.79 & -21 14 18.8 &  30.9 $\pm$  1.3 & 15/06/1998 &     ---    &  35 \\
EISD56  & NVSS-J094728-2128 & 09 47 28.89 & -21 28 35.8 &  28.5 $\pm$  2.9 & 15/06/1998 & 30/04/2001 &  28 \\
EISD57  & NVSS-J095816-2018 & 09 58 16.61 & -20 18 54.3 &  29.6 $\pm$  1.3 & 15/06/1998 &     ---    &  42 \\
EISD58  & NVSS-J094918-2054 & 09 49 18.12 & -20 54 52.5 &  27.5 $\pm$  1.7 & 15/06/1998 &     ---    &  45 \\
EISD59  & NVSS-J094453-2046 & 09 44 53.65 & -20 46 35.9 &  28.6 $\pm$  1.0 & 15/06/1998 &     ---    &  39 \\
EISD60  & NVSS-J095201-2115 & 09 52 01.84 & -21 15 50.7 &  26.5 $\pm$  0.9 & 15/06/1998 &     ---    &  29 \\
EISD61  & NVSS-J095940-2034 & 09 59 40.85 & -20 34 48.3 &  26.3 $\pm$  1.2 & 15/06/1998 &     ---    &  40 \\
\hline
\end{tabular}
\end{table*}
\addtocounter{table}{-1}

\begin{table*}
\caption{{\it continued.} Details of the EIS--NVSS sample of radio sources.}
\begin{tabular}{lcccrccr}
\hline
\multicolumn{1}{c}{Source} &
IAU Name &
\multicolumn{2}{c}{RA~~~(NVSS)~~~Dec}      &
\multicolumn{1}{c}{$S_{\rm 1.4\,GHz}^{\rm NVSS}$} &
\multicolumn{2}{c}{Observation Date} &
\multicolumn{1}{c}{rms} \\
&
&
\multicolumn{2}{c}{(J2000)} &
\multicolumn{1}{c}{[mJy]} &
BnA array &
CnB array &
\multicolumn{1}{c}{[$\mu$Jy]} \\
\hline
EISD62  & NVSS-J095427-2029 & 09 54 27.08 & -20 29 46.7 &  26.1 $\pm$  0.9 & 15/06/1998 &     ---    &  28 \\
EISD63  & NVSS-J094703-2050 & 09 47 03.48 & -20 50 02.0 &  25.2 $\pm$  0.9 & 15/06/1998 &     ---    &  33 \\
EISD64  & NVSS-J095259-2148 & 09 52 59.24 & -21 48 41.4 &  26.4 $\pm$  0.9 & 15/06/1998 &     ---    &  34 \\
EISD65  & NVSS-J095403-2025 & 09 54 03.09 & -20 25 13.3 &  25.2 $\pm$  0.9 & 15/06/1998 &     ---    &  31 \\
EISD66  & NVSS-J095743-2006 & 09 57 43.07 & -20 06 36.1 &  25.5 $\pm$  1.2 & 15/06/1998 &     ---    &  46 \\
EISD67  & NVSS-J095323-2013 & 09 53 23.15 & -20 13 43.6 &  23.8 $\pm$  0.9 & 15/06/1998 &     ---    &  26 \\
EISD68  & NVSS-J095428-2039 & 09 54 28.16 & -20 39 28.2 &  24.2 $\pm$  0.9 & 15/06/1998 &     ---    &  46 \\
EISD69  & NVSS-J095212-2102 & 09 52 12.79 & -21 02 36.2 &  22.3 $\pm$  0.8 & 15/06/1998 &     ---    &  36 \\
EISD70  & NVSS-J095130-2204 & 09 51 30.71 & -22 04 27.7 &  22.3 $\pm$  1.1 & 15/06/1998 &     ---    &  30 \\
EISD71  & NVSS-J094930-2023 & 09 49 30.68 & -20 23 33.4 &  21.4 $\pm$  0.8 & 15/06/1998 &     ---    &  27 \\
EISD72  & NVSS-J094542-2115 & 09 45 42.59 & -21 15 42.3 &  21.7 $\pm$  0.8 & 15/06/1998 &     ---    &  41 \\
EISD73  & NVSS-J095546-2126 & 09 55 46.10 & -21 26 55.6 &  18.9 $\pm$  3.3 & 15/06/1998 & 30/04/2001 &  20 \\
EISD74  & NVSS-J095320-2143 & 09 53 20.62 & -21 43 58.7 &  21.4 $\pm$  0.8 & 15/06/1998 &     ---    &  53 \\
EISD75  & NVSS-J095122-2151 & 09 51 22.83 & -21 51 52.0 &  21.7 $\pm$  0.8 & 15/06/1998 &     ---    &  32 \\
EISD76  & NVSS-J095132-2100 & 09 51 32.53 & -21 00 27.2 &  21.6 $\pm$  1.1 & 15/06/1998 &     ---    &  36 \\
EISD77  & NVSS-J095523-2128 & 09 55 23.78 & -21 28 30.5 &  19.6 $\pm$  2.3 & 15/06/1998 & 30/04/2001 &  18 \\
EISD78  & NVSS-J095043-2126 & 09 50 43.23 & -21 26 37.6 &  20.8 $\pm$  1.1 & 15/06/1998 &     ---    &  34 \\
EISD79  & NVSS-J094855-2103 & 09 48 55.29 & -21 03 57.4 &  20.7 $\pm$  0.8 & 15/06/1998 &     ---    &  29 \\
EISD80  & NVSS-J095121-2129 & 09 51 21.17 & -21 29 54.6 &  20.7 $\pm$  1.1 & 15/06/1998 &     ---    &  36 \\
EISD81  & NVSS-J094842-2152 & 09 48 42.34 & -21 52 26.1 &  19.1 $\pm$  1.1 & 15/06/1998 &     ---    &  34 \\
EISD82  & NVSS-J094801-2009 & 09 48 01.89 & -20 09 11.7 &  18.5 $\pm$  0.7 & 15/06/1998 &     ---    &  37 \\
EISD83  & NVSS-J095148-2031 & 09 51 48.71 & -20 31 53.4 &  18.9 $\pm$  0.7 & 15/06/1998 &     ---    &  28 \\
EISD84  & NVSS-J094945-2150 & 09 49 45.77 & -21 50 04.8 &  18.4 $\pm$  0.7 & 15/06/1998 &     ---    &  35 \\
EISD85  & NVSS-J094859-2050 & 09 48 59.96 & -20 50 07.9 &  18.1 $\pm$  1.0 & 15/06/1998 & 09/06/2002 &  27 \\
EISD86  & NVSS-J095512-2123 & 09 55 12.44 & -21 23 09.6 &  15.3 $\pm$  3.3 & 15/06/1998 &     ---    &  33 \\
EISD87  & NVSS-J095726-2013 & 09 57 26.12 & -20 13 04.3 &  17.9 $\pm$  1.0 & 15/06/1998 &     ---    &  28 \\
EISD88  & NVSS-J094529-2118 & 09 45 29.43 & -21 18 48.8 &  18.3 $\pm$  0.7 & 15/06/1998 &     ---    &  49 \\
EISD89  & NVSS-J095731-2120 & 09 57 31.82 & -21 20 25.4 &  17.3 $\pm$  0.7 & 15/06/1998 &     ---    &  52 \\
EISD90  & NVSS-J095046-2132 & 09 50 46.78 & -21 32 51.1 &  17.4 $\pm$  1.1 & 15/06/1998 &     ---    &  29 \\
EISD91  & NVSS-J095451-2130 & 09 54 51.94 & -21 30 17.5 &  17.2 $\pm$  0.7 & 15/06/1998 &     ---    &  28 \\
EISD92  & NVSS-J095602-2156 & 09 56 02.35 & -21 56 02.9 &  17.0 $\pm$  0.7 & 15/06/1998 &     ---    &  29 \\
EISD93  & NVSS-J095541-2039 & 09 55 41.86 & -20 39 39.3 &  16.7 $\pm$  0.7 & 15/06/1998 &     ---    &  37 \\
EISD94  & NVSS-J095628-2048 & 09 56 28.15 & -20 48 44.5 &  16.2 $\pm$  0.7 & 15/06/1998 &     ---    &  40 \\
EISD95  & NVSS-J094428-2038 & 09 44 28.80 & -20 38 00.3 &  16.2 $\pm$  1.0 & 15/06/1998 &     ---    &  43 \\
EISD96  & NVSS-J094929-2129 & 09 49 29.71 & -21 29 38.8 &  16.0 $\pm$  0.7 & 15/06/1998 &     ---    &  28 \\
EISD97  & NVSS-J094925-2037 & 09 49 25.95 & -20 37 24.5 &  16.5 $\pm$  0.7 & 15/06/1998 &     ---    &  30 \\
EISD98  & NVSS-J094526-2033 & 09 45 26.88 & -20 33 52.8 &  15.7 $\pm$  1.0 & 15/06/1998 &     ---    &  43 \\
EISD99  & NVSS-J094446-2050 & 09 44 46.39 & -20 50 00.8 &  15.7 $\pm$  1.1 & 15/06/1998 &     ---    &  45 \\
EISD100 & NVSS-J095843-2051 & 09 58 43.82 & -20 51 30.1 &  15.4 $\pm$  0.7 & 15/06/1998 &     ---    &  35 \\
EISD101 & NVSS-J095844-2031 & 09 58 44.74 & -20 31 15.0 &  14.8 $\pm$  0.7 & 15/06/1998 &     ---    &  27 \\
EISD102 & NVSS-J095746-2123 & 09 57 46.01 & -21 23 26.5 &  15.3 $\pm$  0.7 & 15/06/1998 &     ---    &  37 \\
EISD103 & NVSS-J095545-2125 & 09 55 45.86 & -21 25 27.0 &  13.5 $\pm$  1.2 & 15/06/1998 & 30/04/2001 &  20 \\
EISD104 & NVSS-J094942-2037 & 09 49 42.89 & -20 37 44.5 &  15.0 $\pm$  0.7 & 15/06/1998 &     ---    &  31 \\
EISD105 & NVSS-J095416-2129 & 09 54 16.90 & -21 29 12.4 &  14.5 $\pm$  1.4 & 19/06/1998 & 30/04/2001 &  40 \\
EISD106 & NVSS-J094548-2159 & 09 45 48.34 & -21 59 08.6 &  14.6 $\pm$  1.1 & 15/06/1998 &     ---    &  43 \\
EISD107 & NVSS-J095559-2042 & 09 55 59.27 & -20 42 53.2 &  14.6 $\pm$  0.7 & 15/06/1998 &     ---    &  26 \\
EISD108 & NVSS-J095741-1955 & 09 57 41.91 & -19 55 57.0 &  14.2 $\pm$  1.0 & 15/06/1998 &     ---    &  25 \\
EISD109 & NVSS-J094447-2058 & 09 44 47.38 & -20 58 14.7 &  14.0 $\pm$  0.6 & 15/06/1998 &     ---    &  42 \\
EISD110 & NVSS-J095453-2115 & 09 54 53.16 & -21 15 12.3 &  14.5 $\pm$  0.6 & 15/06/1998 &     ---    &  32 \\
EISD111 & NVSS-J094556-2120 & 09 45 56.28 & -21 20 48.9 &  13.2 $\pm$  0.6 & 15/06/1998 & 30/04/2001 &  15 \\
EISD112 & NVSS-J095523-2130 & 09 55 23.96 & -21 30 02.9 &  13.4 $\pm$  1.0 & 15/06/1998 & 30/04/2001 &  18 \\
EISD113 & NVSS-J095053-2133 & 09 50 53.40 & -21 33 01.4 &  13.6 $\pm$  0.6 & 15/06/1998 &     ---    &  37 \\
EISD114 & NVSS-J094734-2126 & 09 47 34.67 & -21 26 58.5 &  12.8 $\pm$  0.6 & 15/06/1998 & 30/04/2001 &  28 \\
EISD115 & NVSS-J094445-2146 & 09 44 45.29 & -21 46 21.0 &  13.4 $\pm$  0.6 & 15/06/1998 &     ---    &  36 \\
EISD116 & NVSS-J095129-2016 & 09 51 29.60 & -20 16 41.5 &  13.5 $\pm$  0.6 & 15/06/1998 &     ---    &  28 \\
EISD117 & NVSS-J095309-2001 & 09 53 09.44 & -20 01 22.3 &  13.0 $\pm$  1.0 & 15/06/1998 &     ---    &  37 \\
EISD118 & NVSS-J095724-2129 & 09 57 24.26 & -21 29 43.3 &  10.6 $\pm$  1.9 & 15/06/1998 &     ---    &  35 \\
EISD119 & NVSS-J094520-2201 & 09 45 20.94 & -22 01 18.1 &  13.1 $\pm$  0.6 & 15/06/1998 &     ---    &  43 \\
EISD120 & NVSS-J094804-2034 & 09 48 04.23 & -20 34 35.9 &  13.2 $\pm$  0.6 & 15/06/1998 &     ---    &  33 \\
EISD121 & NVSS-J095456-2205 & 09 54 56.93 & -22 05 05.8 &  10.8 $\pm$  1.6 & 15/06/1998 &     ---    &  27 \\
EISD122 & NVSS-J095255-2052 & 09 52 55.86 & -20 52 07.0 &  12.6 $\pm$  1.1 & 15/06/1998 & 30/04/2001 &  18 \\
EISD123 & NVSS-J095421-2148 & 09 54 21.85 & -21 48 04.1 &  12.2 $\pm$  1.2 & 15/06/1998 &     ---    &  37 \\
EISD124 & NVSS-J094810-2001 & 09 48 10.05 & -20 01 58.1 &  12.1 $\pm$  1.4 & 15/06/1998 & 30/04/2001 &  18 \\
\hline
\end{tabular}
\end{table*}
\addtocounter{table}{-1}

\begin{table*}
\caption{{\it continued.} Details of the EIS--NVSS sample of radio sources.}
\begin{tabular}{lcccrccr}
\hline
\multicolumn{1}{c}{Source} &
IAU Name &
\multicolumn{2}{c}{RA~~~(NVSS)~~~Dec}      &
\multicolumn{1}{c}{$S_{\rm 1.4\,GHz}^{\rm NVSS}$} &
\multicolumn{2}{c}{Observation Date} &
\multicolumn{1}{c}{rms} \\
&
&
\multicolumn{2}{c}{(J2000)} &
\multicolumn{1}{c}{[mJy]} &
BnA array &
CnB array &
\multicolumn{1}{c}{[$\mu$Jy]} \\
\hline
EISD125 & NVSS-J094521-2043 & 09 45 21.17 & -20 43 18.3 &  12.2 $\pm$  0.6 & 15/06/1998 &     ---    &  42 \\
EISD126 & NVSS-J095436-2144 & 09 54 36.25 & -21 44 37.9 &  12.0 $\pm$  1.2 & 15/06/1998 &     ---    &  42 \\
EISD127 & NVSS-J094822-2105 & 09 48 22.17 & -21 05 08.4 &  12.7 $\pm$  0.6 & 15/06/1998 &     ---    &  36 \\
EISD128 & NVSS-J095816-2058 & 09 58 16.47 & -20 58 23.7 &  10.5 $\pm$  1.3 & 15/06/1998 &     ---    &  32 \\
EISD129 & NVSS-J095620-2203 & 09 56 20.42 & -22 03 47.7 &  11.3 $\pm$  0.6 & 15/06/1998 &     ---    &  40 \\
EISD130 & NVSS-J094935-2158 & 09 49 35.26 & -21 58 08.4 &  11.8 $\pm$  0.6 & 15/06/1998 &     ---    &  42 \\
EISD131 & NVSS-J094925-2005 & 09 49 25.83 & -20 05 18.4 &  12.0 $\pm$  0.6 & 15/06/1998 &     ---    &  36 \\
EISD132 & NVSS-J094618-2037 & 09 46 18.92 & -20 37 58.4 &  12.2 $\pm$  0.6 & 15/06/1998 &     ---    &  44 \\
EISD133 & NVSS-J095702-2156 & 09 57 02.06 & -21 56 50.7 &  11.6 $\pm$  0.6 & 15/06/1998 &     ---    &  33 \\
EISD134 & NVSS-J094649-2116 & 09 46 49.75 & -21 16 47.0 &  11.1 $\pm$  1.1 & 15/06/1998 & 30/04/2001 &  13 \\
EISD135 & NVSS-J095851-2110 & 09 58 51.31 & -21 10 20.4 &  11.4 $\pm$  1.1 & 15/06/1998 &     ---    &  38 \\
EISD136 & NVSS-J095048-2154 & 09 50 48.49 & -21 54 55.0 &  11.5 $\pm$  0.6 & 15/06/1998 &     ---    &  40 \\
EISD137 & NVSS-J094813-1959 & 09 48 13.10 & -19 59 54.7 &  10.0 $\pm$  1.2 & 15/06/1998 & 30/04/2001 &  18 \\
EISD138 & NVSS-J094724-2105 & 09 47 24.55 & -21 05 05.6 &  10.6 $\pm$  0.6 & 15/06/1998 &     ---    &  39 \\
EISD139 & NVSS-J095250-2131 & 09 52 50.33 & -21 31 47.3 &  11.4 $\pm$  0.6 & 15/06/1998 &     ---    &  39 \\
EISD140 & NVSS-J095933-2114 & 09 59 33.68 & -21 14 50.8 &   8.9 $\pm$  1.6 & 15/06/1998 &     ---    &  40 \\
EISD141 & NVSS-J095511-2030 & 09 55 11.87 & -20 30 20.6 &  10.1 $\pm$  1.3 & 15/06/1998 & 30/04/2001 &  25 \\
EISD142 & NVSS-J095607-2005 & 09 56 07.01 & -20 05 40.6 &  10.5 $\pm$  0.6 & 15/06/1998 &     ---    &  36 \\
EISD143 & NVSS-J095735-2029 & 09 57 35.04 & -20 29 31.7 &   9.6 $\pm$  0.6 & 15/06/1998 &     ---    &  36 \\
EISD144 & NVSS-J094450-2017 & 09 44 50.70 & -20 17 35.2 &  10.5 $\pm$  0.6 & 15/06/1998 &     ---    &  60 \\
EISD145 & NVSS-J095739-2003 & 09 57 39.46 & -20 03 18.5 &  10.7 $\pm$  0.6 & 15/06/1998 &     ---    &  28 \\
EISD146 & NVSS-J095642-2119 & 09 56 42.22 & -21 19 41.8 &   9.8 $\pm$  0.6 & 15/06/1998 &     ---    &  39 \\
EISD147 & NVSS-J095857-2034 & 09 58 57.31 & -20 34 19.2 &   8.1 $\pm$  1.3 & 15/06/1998 &     ---    &  38 \\
EISD148 & NVSS-J094538-2111 & 09 45 38.34 & -21 11 12.1 &  10.3 $\pm$  1.0 & 15/06/1998 & 09/06/2002 &  30 \\
EISD149 & NVSS-J094744-2112 & 09 47 44.72 & -21 12 29.8 &  10.0 $\pm$  0.6 & 15/06/1998 &     ---    &  48 \\
EISD150 & NVSS-J094710-2035 & 09 47 10.23 & -20 35 53.0 &   9.7 $\pm$  0.6 & 15/06/1998 &     ---    &  53 \\
EISD151 & NVSS-J095553-2127 & 09 55 53.71 & -21 27 12.1 &   4.5 $\pm$  0.8 & 15/06/1998 & 30/04/2001 &  20 \\
EISD152 & NVSS-J095927-2003 & 09 59 27.65 & -20 03 14.8 &   9.7 $\pm$  0.6 & 15/06/1998 &     ---    &  43 \\
EISD153 & NVSS-J095649-2035 & 09 56 49.68 & -20 35 23.9 &  10.2 $\pm$  0.6 & 15/06/1998 &     ---    &  36 \\
EISD154 & NVSS-J095210-2050 & 09 52 10.86 & -20 50 08.2 &  10.1 $\pm$  0.6 & 15/06/1998 &     ---    &  52 \\
EISD155 & NVSS-J095724-2022 & 09 57 24.53 & -20 22 36.3 &   9.6 $\pm$  1.0 & 15/06/1998 &     ---    &  37 \\
EISD156 & NVSS-J095637-2019 & 09 56 37.01 & -20 19 02.6 &   9.0 $\pm$  0.6 & 15/06/1998 &     ---    &  37 \\
EISD157 & NVSS-J094902-2115 & 09 49 02.13 & -21 15 04.7 &   9.4 $\pm$  0.6 & 15/06/1998 &     ---    &  42 \\
EISD158 & NVSS-J094443-2144 & 09 44 43.88 & -21 44 38.5 &   9.3 $\pm$  0.6 & 15/06/1998 &     ---    &  60 \\
EISD159 & NVSS-J095357-2036 & 09 53 57.45 & -20 36 52.0 &   9.1 $\pm$  0.6 & 15/06/1998 &     ---    &  34 \\
EISD160 & NVSS-J095849-2117 & 09 58 49.60 & -21 17 35.5 &   8.8 $\pm$  0.6 & 19/06/1998 &     ---    &  29 \\
EISD161 & NVSS-J094748-2048 & 09 47 48.46 & -20 48 35.7 &   9.4 $\pm$  0.6 & 15/06/1998 &     ---    &  39 \\
EISD162 & NVSS-J095638-2010 & 09 56 38.86 & -20 10 44.1 &   4.1 $\pm$  0.8 & 19/06/1998 & 09/06/2002 &  30 \\
EISD163 & NVSS-J094910-2021 & 09 49 10.67 & -20 21 52.5 &   8.7 $\pm$  0.6 & 19/06/1998 &     ---    &  33 \\
EISD164 & NVSS-J095201-2024 & 09 52 01.26 & -20 24 54.0 &   9.0 $\pm$  0.5 & 19/06/1998 &     ---    &  39 \\
EISD165 & NVSS-J095410-2158 & 09 54 10.47 & -21 58 00.6 &   9.5 $\pm$  0.6 & 19/06/1998 &     ---    &  45 \\
EISD166 & NVSS-J095604-2144 & 09 56 04.41 & -21 44 36.3 &   9.6 $\pm$  0.6 & 19/06/1998 &     ---    &  41 \\
EISD167 & NVSS-J094602-2151 & 09 46 02.57 & -21 51 40.4 &   7.9 $\pm$  0.6 & 19/06/1998 &     ---    &  20 \\
EISD168 & NVSS-J094450-2007 & 09 44 50.35 & -20 07 35.5 &   8.4 $\pm$  0.6 & 19/06/1998 &     ---    &  48 \\
EISD169 & NVSS-J095148-2133 & 09 51 48.92 & -21 33 37.2 &   8.2 $\pm$  0.6 & 19/06/1998 &     ---    &  58 \\
EISD170 & NVSS-J095226-2001 & 09 52 26.52 & -20 01 05.2 &   8.3 $\pm$  0.6 & 19/06/1998 &     ---    &  46 \\
EISD171 & NVSS-J094750-2142 & 09 47 50.04 & -21 42 17.2 &   8.4 $\pm$  1.3 & 19/06/1998 &     ---    &  43 \\
EISD172 & NVSS-J095722-2101 & 09 57 22.22 & -21 01 03.4 &   8.2 $\pm$  0.5 & 19/06/1998 &     ---    &  35 \\
EISD173 & NVSS-J095431-2035 & 09 54 31.12 & -20 35 40.0 &   8.7 $\pm$  0.5 & 19/06/1998 &     ---    &  40 \\
EISD174 & NVSS-J094902-2016 & 09 49 02.62 & -20 16 09.8 &   8.3 $\pm$  0.5 & 19/06/1998 &     ---    &  46 \\
EISD175 & NVSS-J094922-2118 & 09 49 22.34 & -21 18 19.5 &   8.4 $\pm$  0.5 & 19/06/1998 &     ---    &  45 \\
EISD176 & NVSS-J095820-2139 & 09 58 20.40 & -21 39 13.3 &   7.5 $\pm$  1.2 & 19/06/1998 &     ---    &  58 \\
EISD177 & NVSS-J095527-2046 & 09 55 27.05 & -20 46 07.9 &   7.1 $\pm$  0.6 & 19/06/1998 &     ---    &  43 \\
EISD178 & NVSS-J094748-2100 & 09 47 48.27 & -21 00 44.9 &   7.8 $\pm$  0.6 & 19/06/1998 &     ---    &  37 \\
EISD179 & NVSS-J094959-2127 & 09 49 59.68 & -21 27 19.0 &   6.0 $\pm$  0.6 & 19/06/1998 &     ---    &  48 \\
EISD180 & NVSS-J094912-2200 & 09 49 12.22 & -22 00 29.2 &   6.9 $\pm$  0.6 & 19/06/1998 &     ---    &  53 \\
EISD181 & NVSS-J095441-2049 & 09 54 41.98 & -20 49 47.1 &   7.5 $\pm$  0.6 & 19/06/1998 &     ---    &  33 \\
EISD182 & NVSS-J094949-2134 & 09 49 49.04 & -21 34 32.8 &   7.8 $\pm$  0.6 & 19/06/1998 &     ---    &  40 \\
EISD183 & NVSS-J095129-2025 & 09 51 29.20 & -20 25 40.9 &   7.8 $\pm$  1.2 & 19/06/1998 & 30/04/2001 &  14 \\
EISD184 & NVSS-J094507-2109 & 09 45 07.57 & -21 09 28.0 &   6.4 $\pm$  0.6 & 19/06/1998 &     ---    &  52 \\
EISD185 & NVSS-J095212-2140 & 09 52 12.69 & -21 40 27.6 &   4.0 $\pm$  0.7 & 19/06/1998 & 30/04/2001 &  18 \\
EISD186 & NVSS-J094924-2111 & 09 49 24.70 & -21 11 10.0 &   8.3 $\pm$  0.5 & 19/06/1998 &     ---    &  52 \\
EISD187 & NVSS-J095038-2141 & 09 50 38.48 & -21 41 15.1 &   7.4 $\pm$  1.2 & 19/06/1998 & 30/04/2001 &  23 \\
\hline
\end{tabular}
\end{table*}
\addtocounter{table}{-1}

\begin{table*}
\caption{{\it continued.} Details of the EIS--NVSS sample of radio sources.}
\begin{tabular}{lcccrccr}
\hline
\multicolumn{1}{c}{Source} &
IAU Name &
\multicolumn{2}{c}{RA~~~(NVSS)~~~Dec}      &
\multicolumn{1}{c}{$S_{\rm 1.4\,GHz}^{\rm NVSS}$} &
\multicolumn{2}{c}{Observation Date} &
\multicolumn{1}{c}{rms} \\
&
&
\multicolumn{2}{c}{(J2000)} &
\multicolumn{1}{c}{[mJy]} &
BnA array &
CnB array &
\multicolumn{1}{c}{[$\mu$Jy]} \\
\hline
EISD188 & NVSS-J094746-2127 & 09 47 46.11 & -21 27 44.3 &   6.1 $\pm$  0.6 & 19/06/1998 &     ---    &  47 \\
EISD189 & NVSS-J094552-2014 & 09 45 52.01 & -20 14 41.8 &   6.6 $\pm$  0.6 & 19/06/1998 &     ---    &  60 \\
EISD190 & NVSS-J095937-2038 & 09 59 37.53 & -20 38 21.9 &   7.4 $\pm$  0.5 & 19/06/1998 &     ---    &  44 \\
EISD191 & NVSS-J095027-2148 & 09 50 27.69 & -21 48 09.2 &   5.4 $\pm$  0.6 & 19/06/1998 &     ---    &  41 \\
EISD192 & NVSS-J095912-2012 & 09 59 12.27 & -20 12 54.4 &   5.4 $\pm$  0.6 & 19/06/1998 &     ---    &  49 \\
EISD193 & NVSS-J095056-1955 & 09 50 56.59 & -19 55 08.3 &   5.9 $\pm$  0.6 & 19/06/1998 &     ---    &  36 \\
EISD194 & NVSS-J094527-2057 & 09 45 27.50 & -20 57 47.8 &   3.8 $\pm$  0.7 & 19/06/1998 & 30/04/2001 &  17 \\
EISD195 & NVSS-J095715-2030 & 09 57 15.62 & -20 30 34.0 &   6.3 $\pm$  0.6 & 19/06/1998 &     ---    &  42 \\
EISD196 & NVSS-J095002-2205 & 09 50 02.94 & -22 05 34.9 &   5.1 $\pm$  0.6 & 19/06/1998 &     ---    &  57 \\
EISD197 & NVSS-J094522-2036 & 09 45 22.09 & -20 36 12.9 &   4.2 $\pm$  0.7 & 19/06/1998 &     ---    &  57 \\
EISD198 & NVSS-J095850-2108 & 09 58 50.83 & -21 08 14.9 &   6.1 $\pm$  0.6 & 19/06/1998 &     ---    &  43 \\
EISD199 & NVSS-J094526-2154 & 09 45 26.51 & -21 54 57.3 &   6.8 $\pm$  0.5 & 19/06/1998 &     ---    &  43 \\
\hline
\end{tabular}
\end{table*}


\begin{figure*}
\caption{\label{optovers} {\it See attached jpeg files}. Radio maps (in
contours) of those radio sources within the EIS Patch D, with greyscale
representations of the EIS $I$--band image overlaid. The radio contour levels
are at ($-1$,1,2,4,8,16,32,64,128,256,512,1024) $\times$ 3$\sigma$, where
$\sigma$ is the rms noise on the radio map, given in Table~\ref{obsdettab};
any negative contour is shown as a dashed line. The label above each figure
indicates which source is being shown. All figures are shown with an
equivalent grey-scale level, except for EISD162 where this is changed in order
to allow the radio source host galaxy to be distinguished from the nearby
bright star, and the EISD123, EISD163 and EISD191 where the contrast is
lowered to show details of the host galaxy. Likely optical host galaxies are
labelled with crosshairs.}
\end{figure*}

\begin{figure*}
\caption{\label{outovers} {\it See attached jpeg files}. Radio maps (in
contours) of those radio sources outside the EIS Patch D. The radio contour
levels are at ($-1$,1,2,4,8,16,32,64,128,256,512,1024) $\times$ 3$\sigma$,
where $\sigma$ is the rms noise on the radio map, given in
Table~\ref{obsdettab}; any negative contour is shown as a dashed line. The
label above each figure indicates which source is being shown.}
\end{figure*}



\begin{figure*}
\caption{\label{extrasfig} {\it See attached jpeg files}. Radio maps (in
contours) of the three additional radio sources to be added to the CENSORS
sample (CENSORS-X1,X2 and X3), with greyscale representations of the EIS
$I$--band images overlaid. Also, the NVSS map of the potential large double to
be included in the sample (known as CENSORS-D1). The radio contour levels are
at ($-1$,1,2,4,8,16,32,64,128,256,512,1024) $\times$
[100$\mu$Jy,130$\mu$Jy,135$\mu$Jy, 1mJy] for CENSORS-X1,X2,X3 and D1
respectively. Note that all of the CENSOR-X1-3 sources show unphysical
elongations due to bandwidth smearing, since they are observed considerably
off--axis in the radio observations (but they have been corrected for primary
beam attenuation effects).}
\end{figure*}

\clearpage

\begin{footnotesize}
\begin{table*}
\caption{\label{radsourprop} Properties of the radio sources. For each of the
sources, the nature of the source (S=single, D=double, T=triple, M=multiple,
E=extended diffuse), the integrated and peak flux densities in the new radio
observations, the position angle (PA) and largest angular size ($D_{\rm rad}$)
of the radio source, and the location and flux of the radio components are all
listed. In addition, for each source a ``radio source position'' is given; for
extended sources, this is the location of the radio core, where one is
unambiguously detected, or the flux--weighted mean position of the radio
emission where no clear core exists. Notes are provided in the text for 
sources marked with an `N'.}

\begin{tabular}{lcrrrrrllllc}
\hline
Name                                     & 
Morph.                                   & 
\multicolumn{1}{c}{$S_{\rm int}$}     & 
PA                                     & 
$D_{\rm rad}$                                 & 
\multicolumn{1}{c}{$S_{\rm peak}$}    & 
\multicolumn{1}{c}{$S_{\rm comp}$}    & 
\multicolumn{2}{c}{$\alpha_{\rm comp}$~~~(J2000)~~~$\delta_{\rm comp}$}  & 
\multicolumn{2}{c}{$\alpha_{\rm sour}$~~~(J2000)~~~$\delta_{\rm sour}$}  &
Notes                                    \\ 

                                                  & 
                                                  & 
\multicolumn{1}{c}{(mJy)}                         &
\multicolumn{1}{c}{(~$\degr$)}                    & 
\multicolumn{1}{r}{($\arcsec$~~)}                  & 
\multicolumn{1}{c}{(mJy/bm)}                      & 
\multicolumn{1}{c}{(mJy)}                         & 
\multicolumn{1}{l}{~(h~~m~~s)~~}                   &
\multicolumn{1}{l}{~~~~($\degr~~~\arcmin~~~\arcsec$)~~} &
\multicolumn{1}{l}{~(h~~m~~s)~~}                   &
\multicolumn{1}{l}{~~~~($\degr~~~\arcmin~~~\arcsec$)~~} &
                                                  \\
\hline
EISD1     &     D &  634.9 $\pm$   1.2 &   113 &        5.0 &  392.3 &  434.4 $\pm$   0.3 &    09 51 29.00 &  $-$20 50 29.7 &    09 51 29.07 &  $-$20 50 30.1  & \\
          &       &                    &       &            &  162.9 &  196.7 $\pm$   0.3 &    09 51 29.33 &  $-$20 50 31.7 &                &                 & \\
EISD2     &     S &  434.7 $\pm$   1.3 &   132 &        0.9 &  427.0 &  432.2 $\pm$   0.4 &    09 46 50.21 &  $-$20 20 44.4 &    09 46 50.21 &  $-$20 20 44.4  & \\
EISD3     &     S &  348.7 $\pm$   1.2 &   130 &        0.7 &  340.0 &  346.4 $\pm$   0.4 &    09 50 31.39 &  $-$21 02 44.8 &    09 50 31.39 &  $-$21 02 44.8  & \\
EISD4     &     M &  221.1 $\pm$   2.5 &   130 &       38.4 &   28.1 &   34.5 $\pm$   0.2 &    09 49 04.82 &  $-$19 56 58.5 &    09 49 05.81 &  $-$19 57 11.2  & \\
\hline
\end{tabular}
\end{table*}

\addtocounter{table}{-1}

\begin{table*}
\caption{{\it continued.} Properties of the radio sources.}
\begin{tabular}{lcrrrrrllllc}
\hline
Name                                     & 
Morph.                                   & 
\multicolumn{1}{c}{$S_{\rm int}$}     & 
PA                                     & 
$D_{\rm rad}$                                 & 
\multicolumn{1}{c}{$S_{\rm peak}$}    & 
\multicolumn{1}{c}{$S_{\rm comp}$}    & 
\multicolumn{2}{c}{$\alpha_{\rm comp}$~~~(J2000)~~~$\delta_{\rm comp}$}  & 
\multicolumn{2}{c}{$\alpha_{\rm sour}$~~~(J2000)~~~$\delta_{\rm sour}$}  &
Notes                                    \\ 

                                                  & 
                                                  & 
\multicolumn{1}{c}{(mJy)}                         &
\multicolumn{1}{c}{(~$\degr$)}                    & 
\multicolumn{1}{r}{($\arcsec$~~)}                  & 
\multicolumn{1}{c}{(mJy/bm)}                      & 
\multicolumn{1}{c}{(mJy)}                         & 
\multicolumn{1}{l}{~(h~~m~~s)~~}                   &
\multicolumn{1}{l}{~~~~($\degr~~~\arcmin~~~\arcsec$)~~} &
\multicolumn{1}{l}{~(h~~m~~s)~~}                   &
\multicolumn{1}{l}{~~~~($\degr~~~\arcmin~~~\arcsec$)~~} &
                                                  \\
\hline
EISD31    &     D &   44.5 $\pm$   1.2 &    11 &       21.7 &   21.5 &   27.4 $\pm$   0.4 &    09 56 29.90 &  $-$20 01 39.0 &    09 56 30.01 &  $-$20 01 31.0  & \\
          &       &                    &       &            &    9.5 &   16.4 $\pm$   0.5 &    09 56 30.19 &  $-$20 01 17.7 &                &                 & \\
EISD32    &     S &   48.2 $\pm$   0.8 &   143 &        1.4 &   43.1 &   48.4 $\pm$   0.3 &    09 54 38.33 &  $-$21 04 25.1 &    09 54 38.33 &  $-$21 04 25.1  & \\
EISD33    &     D &   46.0 $\pm$   1.7 &     5 &       37.0 &    6.0 &   15.2 $\pm$   0.5 &    09 54 33.58 &  $-$22 05 43.1 &    09 54 33.70 &  $-$22 05 22.3  & \\
          &       &                    &       &            &    7.0 &   19.6 $\pm$   0.6 &    09 54 33.79 &  $-$22 05 06.2 &                &                 & \\
\hline
\end{tabular}
\end{table*}
\addtocounter{table}{-1}

\begin{table*}
\caption{{\it continued.} Properties of the radio sources.}
\begin{tabular}{lcrrrrrllllc}
\hline
Name                                     & 
Morph.                                   & 
\multicolumn{1}{c}{$S_{\rm int}$}     & 
PA                                     & 
$D_{\rm rad}$                                 & 
\multicolumn{1}{c}{$S_{\rm peak}$}    & 
\multicolumn{1}{c}{$S_{\rm comp}$}    & 
\multicolumn{2}{c}{$\alpha_{\rm comp}$~~~(J2000)~~~$\delta_{\rm comp}$}  & 
\multicolumn{2}{c}{$\alpha_{\rm sour}$~~~(J2000)~~~$\delta_{\rm sour}$}  &
Notes                                    \\ 

                                                  & 
                                                  & 
\multicolumn{1}{c}{(mJy)}                         &
\multicolumn{1}{c}{(~$\degr$)}                    & 
\multicolumn{1}{r}{($\arcsec$~~)}                  & 
\multicolumn{1}{c}{(mJy/bm)}                      & 
\multicolumn{1}{c}{(mJy)}                         & 
\multicolumn{1}{l}{~(h~~m~~s)~~}                   &
\multicolumn{1}{l}{~~~~($\degr~~~\arcmin~~~\arcsec$)~~} &
\multicolumn{1}{l}{~(h~~m~~s)~~}                   &
\multicolumn{1}{l}{~~~~($\degr~~~\arcmin~~~\arcsec$)~~} &
                                                  \\
\hline
EISD64    &     D &   18.3 $\pm$   1.0 &   153 &        5.7 &    8.5 &   10.9 $\pm$   0.4 &    09 52 59.10 &  $-$21 48 40.3 &    09 52 59.17 &  $-$21 48 42.4  & \\
          &       &                    &       &            &    5.8 &    7.4 $\pm$   0.4 &    09 52 59.28 &  $-$21 48 45.4 &                &                 & \\
EISD65    &     S &   21.2 $\pm$   0.3 &       &   $<$  1.1 &   21.0 &   21.2 $\pm$   0.3 &    09 54 03.02 &  $-$20 25 13.2 &    09 54 03.02 &  $-$20 25 13.2  & \\
EISD66    &     S &   23.6 $\pm$   1.0 &   161 &        6.2 &    7.6 &   23.1 $\pm$   0.5 &    09 57 42.91 &  $-$20 06 36.1 &    09 57 42.91 &  $-$20 06 36.1  & \\
EISD67    &     S &   26.0 $\pm$   0.3 &       &   $<$  1.0 &   25.5 &   26.0 $\pm$   0.3 &    09 53 23.18 &  $-$20 13 43.5 &    09 53 23.18 &  $-$20 13 43.5  & \\
\hline
\end{tabular}
\end{table*}
\addtocounter{table}{-1}

\begin{table*}
\caption{{\it continued.} Properties of the radio sources.}
\begin{tabular}{lcrrrrrllllc}
\hline
Name                                     & 
Morph.                                   & 
\multicolumn{1}{c}{$S_{\rm int}$}     & 
PA                                     & 
$D_{\rm rad}$                                 & 
\multicolumn{1}{c}{$S_{\rm peak}$}    & 
\multicolumn{1}{c}{$S_{\rm comp}$}    & 
\multicolumn{2}{c}{$\alpha_{\rm comp}$~~~(J2000)~~~$\delta_{\rm comp}$}  & 
\multicolumn{2}{c}{$\alpha_{\rm sour}$~~~(J2000)~~~$\delta_{\rm sour}$}  &
Notes                                    \\ 

                                                  & 
                                                  & 
\multicolumn{1}{c}{(mJy)}                         &
\multicolumn{1}{c}{(~$\degr$)}                    & 
\multicolumn{1}{r}{($\arcsec$~~)}                  & 
\multicolumn{1}{c}{(mJy/bm)}                      & 
\multicolumn{1}{c}{(mJy)}                         & 
\multicolumn{1}{l}{~(h~~m~~s)~~}                   &
\multicolumn{1}{l}{~~~~($\degr~~~\arcmin~~~\arcsec$)~~} &
\multicolumn{1}{l}{~(h~~m~~s)~~}                   &
\multicolumn{1}{l}{~~~~($\degr~~~\arcmin~~~\arcsec$)~~} &
                                                  \\
\hline
EISD105   &     D &   16.4 $\pm$   2.1 &   158 &       40.1 &    1.7 &    4.2 $\pm$   0.6 &    09 54 15.95 &  $-$21 28 45.1 &    09 54 16.43 &  $-$21 29 01.6  & \\
          &       &                    &       &            &    1.2 &    3.4 $\pm$   0.6 &    09 54 17.02 &  $-$21 29 22.3 &                &                 & \\
EISD106   &     S &   11.8 $\pm$   0.8 &   154 &        5.3 &    6.3 &    8.7 $\pm$   0.4 &    09 45 48.48 &  $-$21 59 06.1 &    09 45 48.48 &  $-$21 59 06.1  & \\
EISD107   &     D &   16.0 $\pm$   1.1 &   177 &        6.9 &    3.4 &    8.0 $\pm$   0.6 &    09 55 59.22 &  $-$20 42 48.8 &    09 55 59.23 &  $-$20 42 51.6  & \\
          &       &                    &       &            &    3.0 &    5.3 $\pm$   0.4 &    09 55 59.24 &  $-$20 42 55.7 &                &                 & \\
\hline
\end{tabular}
\end{table*}
\addtocounter{table}{-1}

\begin{table*}
\caption{{\it continued.} Properties of the radio sources.}
\begin{tabular}{lcrrrrrllllc}
\hline
Name                                     & 
Morph.                                   & 
\multicolumn{1}{c}{$S_{\rm int}$}     & 
PA                                     & 
$D_{\rm rad}$                                 & 
\multicolumn{1}{c}{$S_{\rm peak}$}    & 
\multicolumn{1}{c}{$S_{\rm comp}$}    & 
\multicolumn{2}{c}{$\alpha_{\rm comp}$~~~(J2000)~~~$\delta_{\rm comp}$}  & 
\multicolumn{2}{c}{$\alpha_{\rm sour}$~~~(J2000)~~~$\delta_{\rm sour}$}  &
Notes                                    \\ 

                                                  & 
                                                  & 
\multicolumn{1}{c}{(mJy)}                         &
\multicolumn{1}{c}{(~$\degr$)}                    & 
\multicolumn{1}{r}{($\arcsec$~~)}                  & 
\multicolumn{1}{c}{(mJy/bm)}                      & 
\multicolumn{1}{c}{(mJy)}                         & 
\multicolumn{1}{l}{~(h~~m~~s)~~}                   &
\multicolumn{1}{l}{~~~~($\degr~~~\arcmin~~~\arcsec$)~~} &
\multicolumn{1}{l}{~(h~~m~~s)~~}                   &
\multicolumn{1}{l}{~~~~($\degr~~~\arcmin~~~\arcsec$)~~} &
                                                  \\
\hline
EISD145   &     D &    9.2 $\pm$   1.3 &   127 &       31.8 &    1.5 &    1.8 $\pm$   0.2 &    09 57 38.30 &  $-$20 03 10.1 &    09 57 39.51 &  $-$20 03 22.6  & \\
          &       &                    &       &            &    2.4 &    3.5 $\pm$   0.4 &    09 57 40.12 &  $-$20 03 29.0 &                &                 & \\
EISD146   &     S &   10.4 $\pm$   0.3 &       &   $<$  1.1 &   10.1 &   10.4 $\pm$   0.2 &    09 56 42.31 &  $-$21 19 44.6 &    09 56 42.31 &  $-$21 19 44.6  & \\
EISD147   &     T &    3.5 $\pm$   0.7 &    16 &        8.6 &    0.5 &    0.9 $\pm$   0.3 &    09 58 57.19 &  $-$20 34 14.3 &    09 58 57.26 &  $-$20 34 10.4  & \\
          &       &                    &       &            &    0.5 &    0.5 $\pm$   0.2 &    09 58 57.26 &  $-$20 34 10.4 &                &                 & \\
\hline
\end{tabular}
\end{table*}
\addtocounter{table}{-1}

\begin{table*}
\caption{{\it continued.} Properties of the radio sources.}
\begin{tabular}{lcrrrrrllllc}
\hline
Name                                     & 
Morph.                                   & 
\multicolumn{1}{c}{$S_{\rm int}$}     & 
PA                                     & 
$D_{\rm rad}$                                 & 
\multicolumn{1}{c}{$S_{\rm peak}$}    & 
\multicolumn{1}{c}{$S_{\rm comp}$}    & 
\multicolumn{2}{c}{$\alpha_{\rm comp}$~~~(J2000)~~~$\delta_{\rm comp}$}  & 
\multicolumn{2}{c}{$\alpha_{\rm sour}$~~~(J2000)~~~$\delta_{\rm sour}$}  &
Notes                                    \\ 

                                                  & 
                                                  & 
\multicolumn{1}{c}{(mJy)}                         &
\multicolumn{1}{c}{(~$\degr$)}                    & 
\multicolumn{1}{r}{($\arcsec$~~)}                  & 
\multicolumn{1}{c}{(mJy/bm)}                      & 
\multicolumn{1}{c}{(mJy)}                         & 
\multicolumn{1}{l}{~(h~~m~~s)~~}                   &
\multicolumn{1}{l}{~~~~($\degr~~~\arcmin~~~\arcsec$)~~} &
\multicolumn{1}{l}{~(h~~m~~s)~~}                   &
\multicolumn{1}{l}{~~~~($\degr~~~\arcmin~~~\arcsec$)~~} &
                                                  \\
\hline
EISD191   &     E &    0.7 $\pm$   0.4 &       &            &    0.5 &    0.6 $\pm$   0.3 &    09 50 27.68 &  $-$21 48 08.7 &    09 50 27.68 &  $-$21 48 08.7  & \\
EISD192   &     S &    4.4 $\pm$   0.3 &       &   $<$  2.1 &    3.8 &    4.4 $\pm$   0.3 &    09 59 12.53 &  $-$20 12 53.2 &    09 59 12.53 &  $-$20 12 53.2  & \\
EISD193   &     S &    3.9 $\pm$   0.4 &       &   $<$  2.5 &    3.7 &    3.9 $\pm$   0.3 &    09 50 55.91 &  $-$19 55 12.0 &    09 50 55.91 &  $-$19 55 12.0  & \\
EISD194   &     D &    1.9 $\pm$   0.4 &    29 &       23.1 &    0.7 &    1.0 $\pm$   0.2 &    09 45 27.39 &  $-$20 57 44.1 &    09 45 27.69 &  $-$20 57 35.2  & \\
          &       &                    &       &            &    0.3 &    0.6 $\pm$   0.2 &    09 45 28.18 &  $-$20 57 23.9 &                &                 & \\
\hline								
\end{tabular}
\end{table*}
\end{footnotesize}

\section{The optical counterparts}
\label{appoptprop}

In this appendix, the technique used to identify optical galaxies in the EIS
catalogue as the host galaxies of the radio sources is described. Then,
Table~\ref{optprops} provides details of the optical counterparts, as
determined from the EIS database.

\subsection{Maximum likelihood analysis}
\label{appmaxlike}

The likelihood ratio technique (e.g. Richter 1975, de Ruiter \etal\
1977)\nocite{ric75,rui77} can be used to statistically investigate whether a
proposed optical identification is the real counterpart of a radio source. In
this method, the dimensionless difference between the radio and optical
positions is defined as:

\begin{displaymath}
r = \left ( \frac{\Delta\alpha^2}{\sigma_{\alpha}^2} +
\frac{\Delta\delta^2}{\sigma_{\delta}^2} \right )^{\frac{1}{2}},
\end{displaymath}

\noindent where $\Delta\alpha$ and $\Delta\delta$ are the differences in right
ascension and declination between the radio and optical positions, and
$\sigma_{\alpha}$ and $\sigma_{\delta}$ are the positional standard errors on
these differences. These have contributions from the uncertainties in both the
radio and optical positions: $\sigma_{\alpha}^2 = \sigma_{\alpha_{\rm rad}}^2
+ \sigma_{\alpha_{\rm opt}}^2$, and $\sigma_{\delta}^2 = \sigma_{\delta_{\rm
rad}}^2 + \sigma_{\delta_{\rm opt}}^2$. The radio uncertainties for a typical
unresolved component are about $\pm 0.3$ arcsec in each direction. Defining
the astrometry of the radio frames to be absolute, the uncertainties in the
optical positions are dominated by the radio--optical astrometric errors,
which are of order $\pm 0.2$ arcsec.

Assuming a random distribution of objects in the EIS survey, the probability
of the nearest random object (c) lying at a radius of between $r$ and $r+{\rm
d}r$ is governed by a Poisson process and can be written as ${\rm d}p(r|{\rm
c}) = 2\lambda r e^{-\lambda r^2} {\rm d}r$, where $\lambda = \pi
\sigma_{\alpha} \sigma_{\delta} \rho$ and $\rho$ is the number of objects per
square arcsecond in the EIS catalogue.  The probability of finding the true
optical counterpart (id) of the radio source at a distance between $r$ and $r
+ {\rm d}r$ can be represented by a Rayleigh distribution: ${\rm d}p(r|{\rm
id}) = r e^{-\lambda r^2 / 2} {\rm d}r$. The likelihood ratio, defined as:

\begin{displaymath}
LR(r) \equiv \frac{{\rm d}p(r|{\rm id})}{{\rm d}p(r|{\rm c})} = \frac{1}{2
\lambda}  e^{\frac{r^2}{2}(2 \lambda - 1)},
\end{displaymath}

\noindent is therefore an effective estimator to distinguish whether
individual optical galaxies are true optical counterparts in the tail of the
Rayleigh distribution or confusing background sources. 

\subsection{Radio sources with well--defined positions}

The likelihood ratio was calculated in this way for every optical galaxy
within 5 arcseconds of any single--component radio source, or any radio source
with a clear radio core. To investigate the appropriate cut--off value of the
likelihood ratio, $LR \ge L$, then (following de Ruiter \etal\
1977)\nocite{rui77} the completeness $C(L)$ (the fraction of real
identifications which are accepted) and the reliability $R(L)$ (the fraction
of accepted identifications which are correct) of the sample can be determined
for different values of $L$ according to the equations:

\begin{displaymath}
C(L) = 1-\frac{1}{N_{\rm id}} \sum_{LR_i < L} p_i({\rm id}|r)
\end{displaymath}

\begin{displaymath}
R(L) = 1-\frac{1}{N_{\rm id}} \sum_{LR_i \ge L} p_i({\rm c}|r) 
\end{displaymath}

\noindent where $p_i(\rm{id}|r)$ and $p_i(\rm{c}|r)$ are the posteriori
probabilities for an object found at a normalised distance $r$ to be the 
true optical counterpart of the radio source or a confusing background object
respectively, and $N_{\rm id}$ is the total number of real identifications
obtained by summing $p_i(\rm{id}|r)$ for all sources. If the a priori
probability of finding an optical counterpart to a source is written as
$p(\rm{id}) = \phi$ then these probabilities can be written as:

\begin{displaymath}
p_i(\rm{id}|r) = \frac{\phi LR_i}{\phi LR_i + 1 - \phi}, \hspace*{0.7cm}
p_i(\rm{c}|r) = \frac{1 - \phi}{\phi LR_i + 1 - \phi}
\end{displaymath}

\noindent and hence, from $\phi$ and the values of $LR_i$, the completeness
and the reliability of the sample can be computed as a function of $L$. 

\begin{figure}
\centerline{
\psfig{file=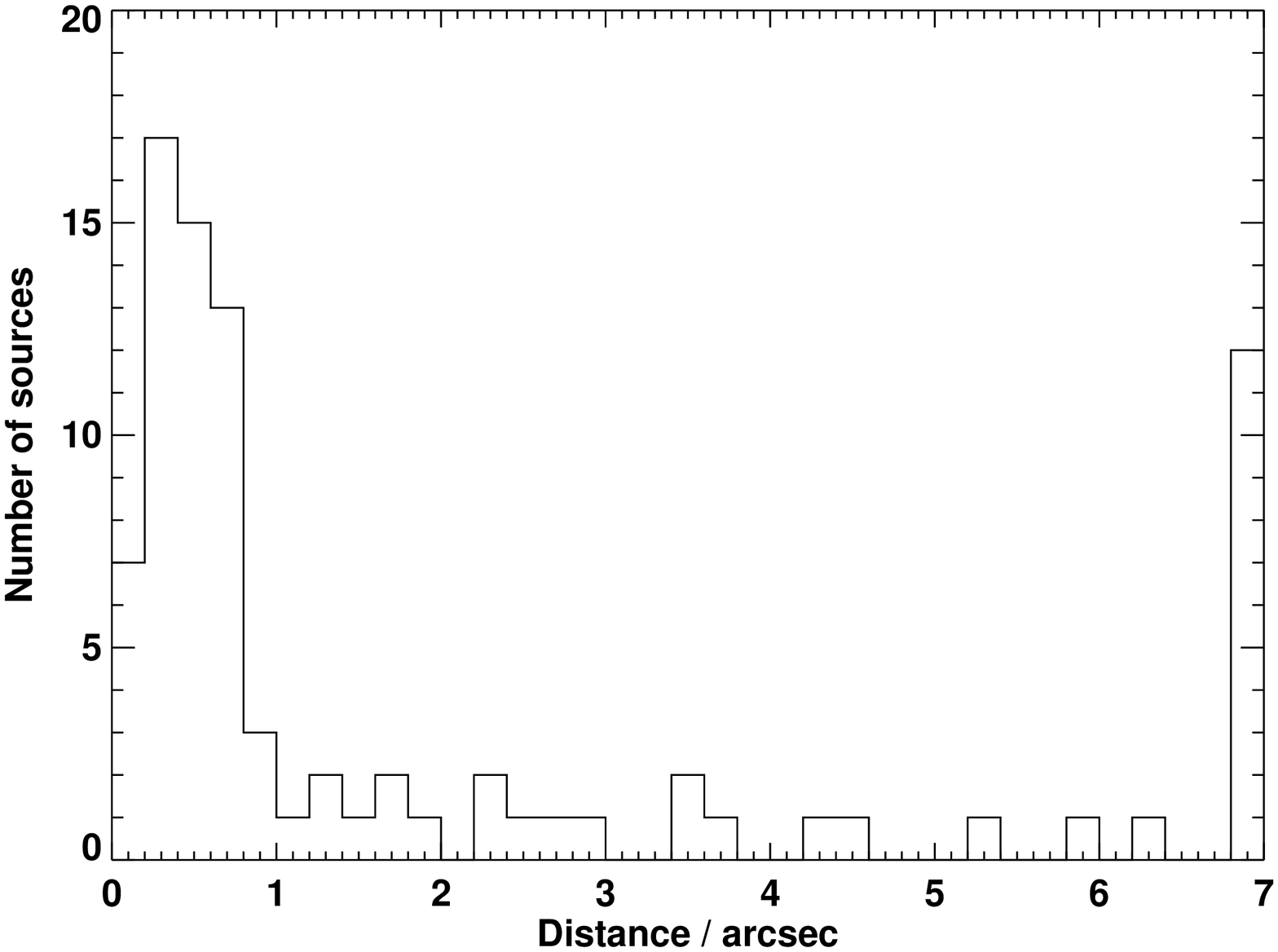,angle=90,width=8.5cm,clip=} 
}
\caption{\label{optcand} A histogram of angular offsets from the radio source
to the nearest object in the EIS catalogue, for all radio sources whose
central position is clearly defined, either as a single--component radio
sources or a source with a clear core. The last bin also contains those
sources for which the nearest object is more than 7 arcsec from the radio
position.}
\end{figure}

To obtain an estimate of $\phi$, for all of the single--component radio
sources or those with a clear core, the angular separation between the radio
position and the nearest optical galaxy was calculated, and the distribution
of these is shown in Figure~\ref{optcand}. There is a clear peak of optical
objects associated with the radio sources at small angular offsets, with the
excess essentially continuing out to about 1 arcsec. 55 of the 87 sources are
found within this radius, and a value of $\phi \approx 55 / 87 = 0.63$ is
adopted.

\begin{figure}
\centerline{
\psfig{file=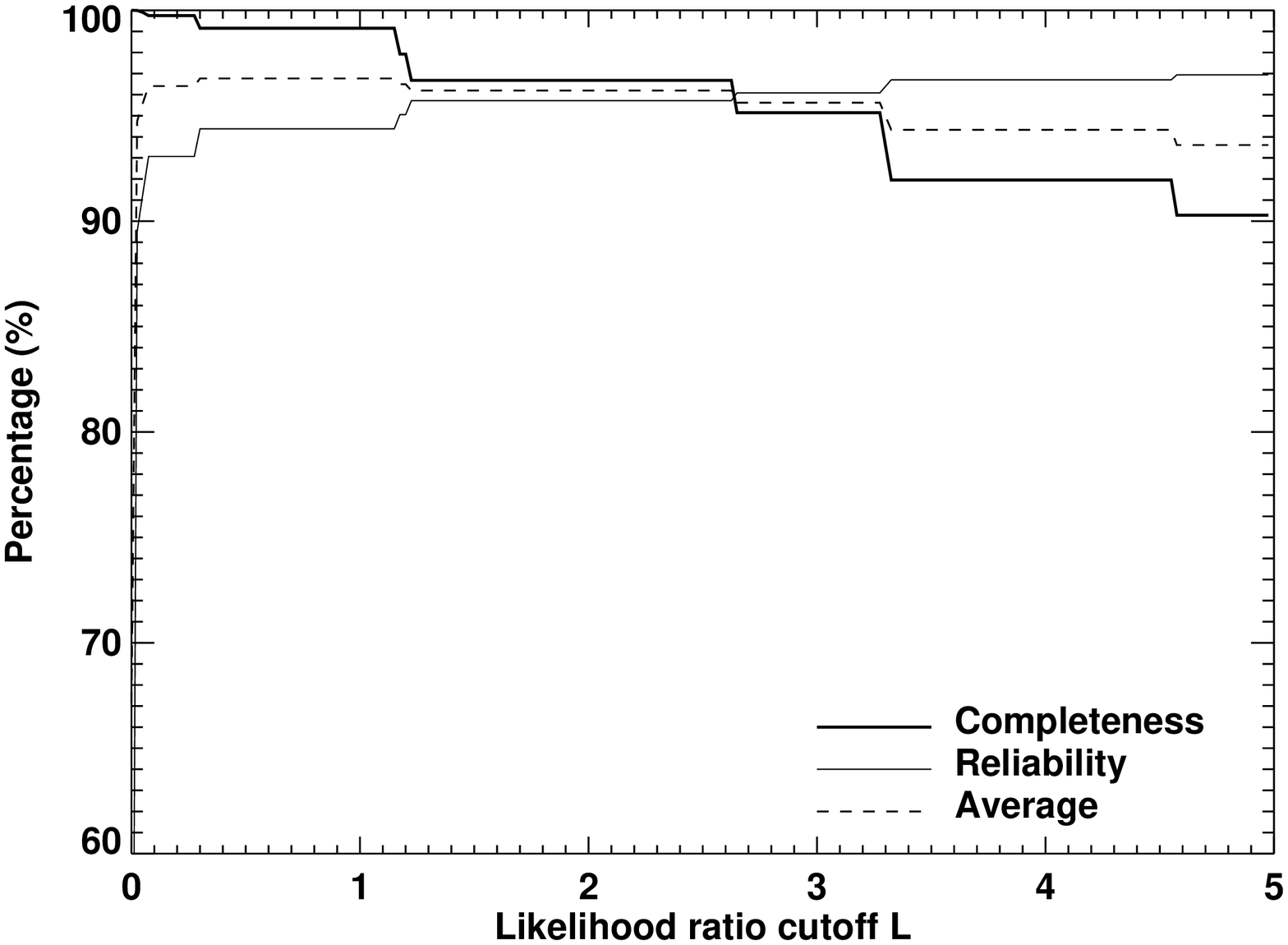,angle=90,width=8.5cm,clip=} 
}
\caption{\label{likept} The completeness and reliability of the optical
identifications as a function of the likelihood ratio cut-off, for all of the
radio sources with a well defined position for their nucleus.} 
\end{figure}

Adopting this value of $\phi$, the completeness and reliability of the
optical identifications are shown as a function of the likelihood ratio
cut--off in Figure~\ref{likept}.  The average of these two functions peaks for
likelihood ratio cut-offs in the range $0.3 \lta L \lta 1.2$; a cut-off value
of $L=1$ is therefore adopted. At this value, the completeness of the optical
identifications should be $\sim 99$\% (meaning that for all radio sources
whose optical host is bright enough to be detected in the EIS-D catalogue,
99\% of them will have been found), and a reliability of over 94\%. 

\subsection{Extended radio sources}

Where no clear radio core is detected for a double, triple, or multiple
component radio source, the source extendness introduces an additional
uncertainty into the radio position which must be correctly accounted for 
in order to accurately estimate the reliability, completeness and
identification fraction of the sample.

Bright radio source samples such as the revised 3CR sample \cite{lai83}, for
which optical identifications have been made for essentially all sources, are
known to frequently have two radio lobes of unequal length, or to show a bend
in their radio jets. These parameters can be quantified as the {\it separation
quotient}, $Q = \theta_1 / \theta_2$, where $\theta_1$ and $\theta_2$ are the
angular distances from the cores to the hotspots of the longer and shorter
lobes respectively, and the {\it asymmetry angle}, $\xi$, that is defined as
180$^{\circ}$ minus the observed angle between vectors drawn from the nucleus
of the radio source to the hotspots. From these two parameters and the size of
the radio source ($D_{\rm rad} = \theta_1 + \theta_2$, assuming that $\xi$ is
small), expressions can be derived for the expected displacement of the
central optical object along and perpendicular to (respectively) the main
radio axis that connects the two hotspots:

\begin{displaymath}
\sigma_Q = \frac{1}{2}(\theta_1 - \theta_2) = \frac{1}{2} D_{\rm rad}
\left(\frac{Q-1}{Q+1} \right ), \hspace*{0.7cm} \sigma_\xi \approx \frac{1}{2} D_{\rm rad} {\rm tan}\frac{\xi}{2}
\end{displaymath}

\noindent assuming that Q is close to unity and $\xi$ is relatively small.

Best \etal\ \shortcite{bes95a} investigated the distribution of $Q$ and $\xi$
within a complete subsample of 95 radio sources from the revised 3CR
sample. They find mean values of $\overline{Q} = 1.42$ and $\overline{\xi} =
6.8^{\circ}$; these mean values were adopted as input for the determination of
the revised positional uncertainties of the extended radio sources in the
sample\footnote{One important point to note with this analysis, however, is
that although Best \etal\ \shortcite{bes95a} find no evidence for any
dependence of $Q$ or $\xi$ upon radio luminosity, the current sample is of
significantly lower luminosity, and is selected at 1.4\,GHz instead of at
178\,MHz. Therefore the distributions of $Q$ and $\xi$ may not be entirely
appropriate, which could potentially bias the statistics of this
investigation.}, according to:

\begin{displaymath}
\sigma_{\alpha_{\rm rad}}^2 = 1''^2 + (\sigma_{Q} {\rm sin PA})^2 +
(\sigma_{\xi} {\rm cos PA})^2
\end{displaymath}

\begin{displaymath}
\sigma_{\delta_{\rm rad}}^2 = 1''^2 + (\sigma_{Q} {\rm cos PA})^2 +
(\sigma_{\xi} {\rm sin PA})^2
\end{displaymath}

\noindent where PA is the position angle of the radio source and the 1.0
arcsecond term corresponds to the uncertainties in the measurement of the
radio source positions; this was the only term used in the analysis of the
single component sources, where a value of 0.3 arcsec was adopted. A larger
value is adopted here to account for the increased uncertainty in measuring
the precise positions of the lobes, since these structures are extended and
faint, but in most cases this term is negligible anyway.

\begin{figure}
\centerline{
\psfig{file=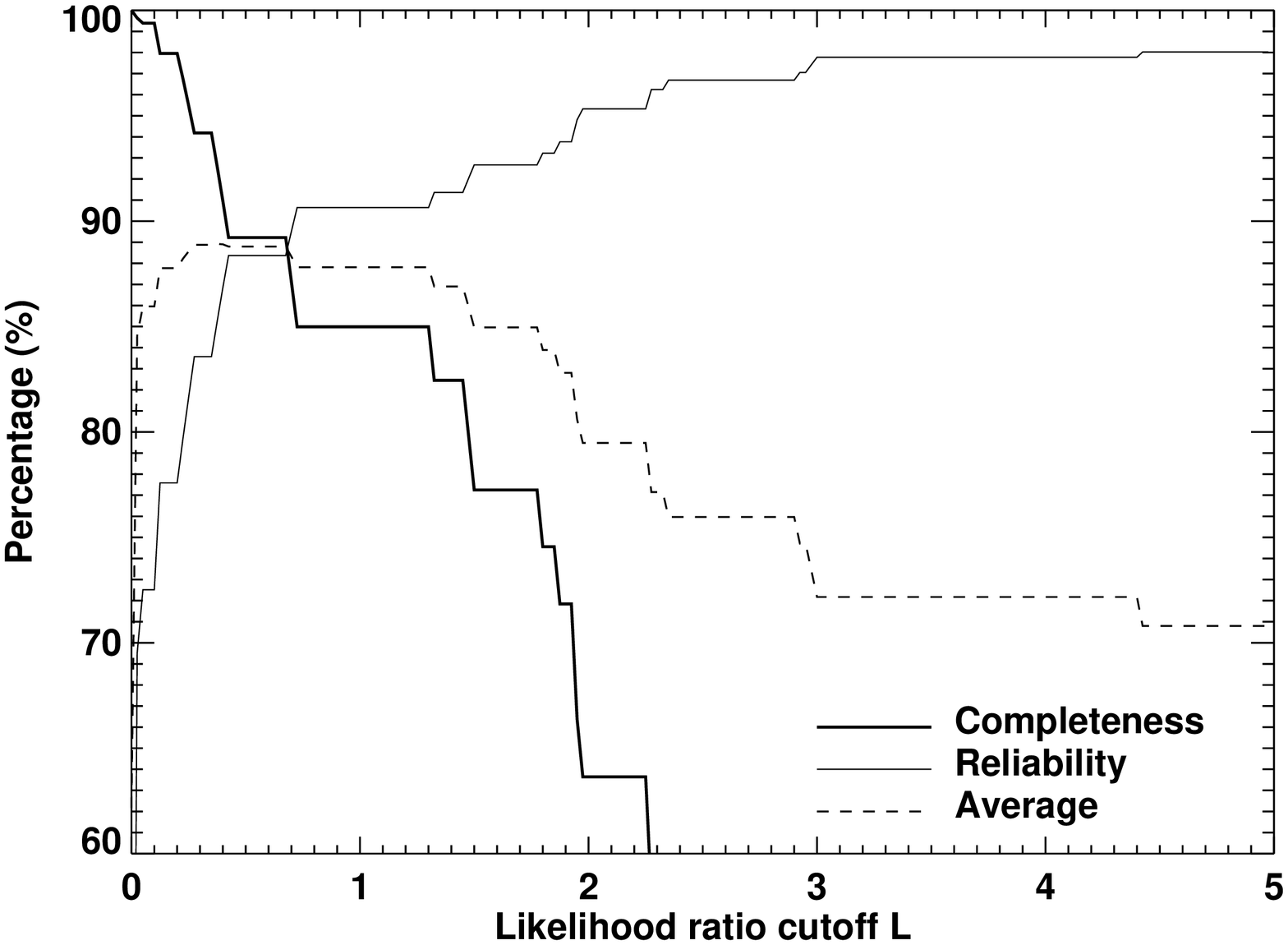,angle=90,width=8.5cm,clip=} 
}
\caption{\label{likeext}  The completeness and reliability of the optical
identifications as a function of the likelihood ratio cut-off, for the
extended radio sources with no well--defined nucleus.} 
\end{figure}

Adopting these values for the radio positional uncertainties of each source,
the likelihood analysis was repeated for the extended radio sources
considering all optical counterparts within 0.25$D_{\rm rad}$ or 10$''$
(whichever was larger) of the position midway between the two radio
hotspots. A value of $\phi = 0.63$ was adopted, as for the single--component
sources; this ignores the possibility that the redshift or magnitude
distribution of the extended sources may be different from that of the
unresolved sources, but there is no better approximation that can be made, and
the overall results vary only slightly if the value of $\phi$ is changed.

The resulting dependence of the completeness and reliability of the identified
optical hosts are shown in Figure~\ref{likeext} as a function of the
likelihood ratio cut--off. The situation is understandably worse than in the
previous case, due to the larger uncertainties in the radio position. At the
peak in the average of the two functions, $L \approx 0.3$, the completeness of
the optical identifications is $\sim 94$\% and the reliability $\sim
84$\%. Optical counterparts above this threshold were found for 34 sources,
with two potential counterparts above the threshold in 2 cases.

It was further noted that for 6 of the sources (see Table~\ref{optprops}) for
which no optical counterpart with a likelihood ratio above the cut--off were
found, clear optical counterparts were seen to be associated with one of the
two apparent radio lobes. These optical galaxies are also accepted as possible
identifications, due to the possibility of radio sources with very asymmetric
flux ratios, where the detected ``lobe'' is actually the radio core and a
fainter lobe on the other side is not detected in the new radio observations.

\begin{table*}
\caption{\label{optprops} EIS optical counterparts of the radio sources. These
were identified by maximum likelihood analysis on unresolved radio sources or
those with unambiguous radio cores (class 1, likelihood $L>1$ for for
inclusion), maximum likelihood analysis on extended sources (class 2, $L>0.3$
for inclusion), or inclusion because the optical counterpart lay directly on
top of one of the radio components which could plausibly be a radio core
(class 3). The optical magnitudes and uncertainties are taken from the EIS
catalogues (see text for details), and the stellaricity classification (S/G;
see text for details) is taken from the EIS $I$--band observations. Notes are
attached in the text for sources marked with an `N'}
\begin{tabular}{lcrrrccccc}
\hline
\multicolumn{1}{c}{Source}  &
\multicolumn{1}{c}{Class}   &
\multicolumn{1}{c}{$L$}     &
\multicolumn{2}{c}{RA~~~~(J2000)~~~~Dec} &
\multicolumn{1}{c}{$I$-magnitude} &
\multicolumn{1}{c}{$V$-magnitude} &
\multicolumn{1}{c}{$B$-magnitude} &
\multicolumn{1}{c}{S/G} &
\multicolumn{1}{c}{Notes} \\
\hline
  EISD1 & 2 & 11.35 & 09 51 29.22 & -20 50 30.7 & 21.74 $\pm$ 0.10 & 22.93 $\pm$ 0.09 & 23.27 $\pm$ 0.09 & 0.26 \\
  EISD2 & 1 & 77.73 & 09 46 50.22 & -20 20 44.4 & 22.59 $\pm$ 0.18 &                  & 24.02 $\pm$ 0.17 & 0.01 \\
  EISD3 & 1 &  5.81 & 09 50 31.40 & -21 02 44.4 & 20.60 $\pm$ 0.20 & 23.11 $\pm$ 0.11 & 23.77 $\pm$ 0.16 & 0.02 & N \\
  EISD6 & 2 &  2.91 & 09 49 53.26 & -21 56 19.9 & 21.30 $\pm$ 0.08 & 23.36 $\pm$ 0.12 & 23.41 $\pm$ 0.13 & 0.00 & N \\
        & 2 &  2.39 & 09 49 53.23 & -21 56 22.0 & 22.58 $\pm$ 0.15 &                  &                  & 0.29 \\
  EISD7 & 1 & 38.80 & 09 51 43.60 & -21 23 57.9 & 18.26 $\pm$ 0.01 & 18.65 $\pm$ 0.01 & 18.40 $\pm$ 0.01 & 1.00 \\
  EISD8 & 1 &  1.22 & 09 53 44.47 & -21 36 01.7 & 22.15 $\pm$ 0.12 & 22.73 $\pm$ 0.06 & 22.57 $\pm$ 0.15 & 0.19 \\
 EISD10 & 1 &  3.31 & 09 45 56.68 & -21 16 53.6 & 22.62 $\pm$ 0.17 & 23.54 $\pm$ 0.19 & 23.38 $\pm$ 0.11 & 0.64 \\
 EISD11 & 1 &  2.65 & 09 57 30.06 & -21 30 58.9 & 17.77 $\pm$ 0.01 & 19.38 $\pm$ 0.02 & 20.35 $\pm$ 0.03 & 0.25 \\
 EISD12 & 1 & 37.18 & 09 49 35.46 & -21 56 23.3 & 18.28 $\pm$ 0.01 & 20.08 $\pm$ 0.01 & 20.96 $\pm$ 0.02 & 0.03 \\
 EISD15 & 1 & 19.77 & 09 53 29.55 & -20 02 12.6 & 21.75 $\pm$ 0.07 & 22.49 $\pm$ 0.05 & 22.74 $\pm$ 0.05 & 0.91 \\
 EISD16 & 2 &  0.31 & 09 47 27.00 & -21 26 33.4 & 18.26 $\pm$ 0.01 & 18.73 $\pm$ 0.01 & 19.15 $\pm$ 0.01 & 0.99 & N \\
 EISD18 & 1 & 24.93 & 09 46 41.13 & -20 29 26.7 & 21.88 $\pm$ 0.09 & 23.35 $\pm$ 0.15 & 23.90 $\pm$ 0.19 & 0.00 \\
 EISD21 & 2 &  6.34 & 09 54 47.59 & -20 59 44.3 & 23.60 $\pm$ 0.25 &                  &                  & 0.58 \\
 EISD22 & 2 &  5.50 & 09 46 50.99 & -20 53 18.2 & 20.57 $\pm$ 0.06 & 20.91 $\pm$ 0.04 & 21.45 $\pm$ 0.03 & 0.00 \\
 EISD24 & 2 &  8.45 & 09 52 43.02 & -19 58 21.6 & 21.38 $\pm$ 0.07 & 23.63 $\pm$ 0.14 &                  & 0.15 \\
 EISD25 & 1 & 12.94 & 09 55 13.59 & -21 23 02.4 & 14.88 $\pm$ 0.00 & 16.13 $\pm$ 0.01 & 16.96 $\pm$ 0.01 & 0.03 \\
 EISD28 & 1 & 41.89 & 09 47 58.95 & -21 21 50.5 & 22.11 $\pm$ 0.10 & 24.38 $\pm$ 0.19 & 24.52 $\pm$ 0.14 & 0.26 \\
 EISD38 & 3 &  N/A  & 09 46 32.15 & -20 26 15.5 & 19.02 $\pm$ 0.02 & 20.86 $\pm$ 0.04 & 21.59 $\pm$ 0.03 & 0.03 & N \\
 EISD39 & 1 & 32.38 & 09 48 15.74 & -21 40 06.1 & 18.73 $\pm$ 0.01 & 19.02 $\pm$ 0.01 & 19.23 $\pm$ 0.01 & 0.97 \\
 EISD40 & 1 &  7.02 & 09 45 55.92 & -20 28 29.7 & 16.41 $\pm$ 0.00 & 18.58 $\pm$ 0.01 & 19.50 $\pm$ 0.05 & 0.86 \\
 EISD43 & 2 &  1.45 & 09 51 40.85 & -20 11 16.3 & 22.25 $\pm$ 0.12 &                  &                  & 0.00 \\
 EISD44 & 1 &  6.79 & 09 51 49.83 & -21 24 57.5 & 18.29 $\pm$ 0.01 & 20.12 $\pm$ 0.02 & 21.14 $\pm$ 0.02 & 0.03 \\
 EISD47 & 1 &  5.78 & 09 47 53.61 & -21 47 19.3 & 24.95 $\pm$ 0.63 &                  &                  & 0.36 & N \\
 EISD48 & 2 & 12.52 & 09 54 52.46 & -21 19 28.6 & 18.71 $\pm$ 0.01 & 21.11 $\pm$ 0.03 & 22.12 $\pm$ 0.04 & 0.03 \\
 EISD51 & 1 & 16.75 & 09 49 33.24 & -21 27 07.7 & 22.90 $\pm$ 0.17 &                  & 24.74 $\pm$ 0.16 & 0.68 \\
 EISD52 & 2 &  1.87 & 09 49 19.57 & -21 51 34.1 & 23.67 $\pm$ 0.19 & 24.55 $\pm$ 0.26 & 24.52 $\pm$ 0.20 & 0.44 \\
 EISD54 & 2 &  1.78 & 09 48 36.08 & -21 06 22.4 & 20.60 $\pm$ 0.03 & 21.73 $\pm$ 0.03 & 21.61 $\pm$ 0.02 & 0.92 \\
 EISD55 & 2 &  2.35 & 09 50 59.01 & -21 14 24.4 & 21.54 $\pm$ 0.10 & 23.18 $\pm$ 0.13 & 23.66 $\pm$ 0.12 & 0.00 \\
 EISD58 & 1 & 15.30 & 09 49 18.16 & -20 54 44.8 & 17.10 $\pm$ 0.01 & 18.92 $\pm$ 0.03 & 20.09 $\pm$ 0.01 & 0.03 \\
 EISD60 & 3 &  N/A  & 09 52 01.59 & -21 15 53.0 & 23.70 $\pm$ 0.19 &                  & 25.06 $\pm$ 0.28 & 0.49 & N\\
 EISD62 & 1 & 47.94 & 09 54 27.09 & -20 29 46.6 & 19.85 $\pm$ 0.02 & 20.48 $\pm$ 0.01 & 20.80 $\pm$ 0.02 & 0.98 \\
 EISD63 & 2 &  0.70 & 09 47 03.25 & -20 50 00.8 & 19.34 $\pm$ 0.02 & 21.08 $\pm$ 0.02 & 21.82 $\pm$ 0.04 & 0.02 \\
 EISD64 & 2 & 11.64 & 09 52 59.15 & -21 48 42.5 & 20.93 $\pm$ 0.05 & 24.37 $\pm$ 0.14 &                  & 0.24 \\
 EISD65 & 1 & 32.03 & 09 54 03.05 & -20 25 13.1 & 20.39 $\pm$ 0.05 & 22.70 $\pm$ 0.10 & 23.53 $\pm$ 0.13 & 0.01 \\
 EISD66 & 1 &  8.08 & 09 57 42.91 & -20 06 36.9 & 20.50 $\pm$ 0.04 & 23.33 $\pm$ 0.10 & 24.16 $\pm$ 0.15 & 0.04 \\
 EISD67 & 1 & 40.69 & 09 53 23.21 & -20 13 43.5 & 19.20 $\pm$ 0.02 & 20.57 $\pm$ 0.16 & 20.55 $\pm$ 0.01 & 0.04 \\
 EISD68 & 1 & 14.13 & 09 54 28.32 & -20 39 26.9 & 19.48 $\pm$ 0.02 & 20.44 $\pm$ 0.01 & 20.41 $\pm$ 0.01 & 0.97 \\
 EISD71 & 2 &  2.25 & 09 49 30.81 & -20 23 34.7 & 19.61 $\pm$ 0.03 & 22.00 $\pm$ 0.18 & 22.94 $\pm$ 0.11 & 0.03 \\
 EISD74 & 1 & 21.23 & 09 53 20.60 & -21 43 58.9 & 15.75 $\pm$ 0.00 & 17.43 $\pm$ 0.01 & 18.59 $\pm$ 0.01 & 0.03 \\
 EISD76 & 2 & 10.61 & 09 51 32.39 & -21 00 29.0 & 17.50 $\pm$ 0.01 & 19.89 $\pm$ 0.02 & 20.95 $\pm$ 0.05 & 0.03 \\
 EISD78 & 2 &  0.35 & 09 50 43.16 & -21 26 41.3 & 22.58 $\pm$ 0.14 &                  &                  & 0.04 \\
 EISD80 & 2 &  7.00 & 09 51 21.03 & -21 29 54.8 & 22.19 $\pm$ 0.13 &                  &                  & 0.14 \\
 EISD81 & 1 &  1.17 & 09 48 42.52 & -21 52 24.8 & 21.69 $\pm$ 0.10 &                  &                  & 0.00 \\
 EISD83 & 1 & 29.40 & 09 51 48.67 & -20 31 52.4 & 23.48 $\pm$ 0.26 &                  &                  & 0.20 \\
 EISD84 & 2 & 09.66 & 09 49 45.86 & -21 50 06.5 & 20.25 $\pm$ 0.03 & 22.94 $\pm$ 0.07 & 24.30 $\pm$ 0.14 & 0.04 \\
 EISD88 & 2 &  7.58 & 09 45 29.53 & -21 18 51.7 & 17.96 $\pm$ 0.01 & 19.96 $\pm$ 0.01 & 20.89 $\pm$ 0.01 & 0.98 \\
 EISD89 & 2 &  1.95 & 09 57 31.77 & -21 20 29.2 & 18.62 $\pm$ 0.01 & 20.87 $\pm$ 0.03 & 21.95 $\pm$ 0.04 & 0.03 \\
 EISD90 & 1 &  2.55 & 09 50 46.42 & -21 32 55.7 & 21.96 $\pm$ 0.11 & 22.47 $\pm$ 0.06 & 23.34 $\pm$ 0.13 & 0.25 \\
 EISD91 & 3 &  N/A  & 09 54 51.95 & -21 30 16.4 & 19.88 $\pm$ 0.20 & 21.54 $\pm$ 0.20 & 22.55 $\pm$ 0.20 & 0.02 & N \\ 
 EISD98 & 2 &  1.49 & 09 45 26.93 & -20 33 52.8 & 16.73 $\pm$ 0.01 & 18.37 $\pm$ 0.01 & 19.88 $\pm$ 0.01 & 0.03 \\
EISD102 & 3 &  N/A  & 09 57 46.06 & -21 23 27.7 & 17.23 $\pm$ 0.01 & 18.99 $\pm$ 0.01 & 20.16 $\pm$ 0.01 & 0.03 & N\\
EISD104 & 1 & 66.46 & 09 49 42.99 & -20 37 45.6 & 23.60 $\pm$ 0.20 &                  & 24.99 $\pm$ 0.31 & 0.53 \\
EISD105 & 2 &  2.98 & 09 54 16.45 & -21 29 04.3 & 21.16 $\pm$ 0.06 & 22.44 $\pm$ 0.08 & 22.87 $\pm$ 0.08 & 0.13 \\
EISD106 & 1 & 48.67 & 09 45 48.50 & -21 59 06.1 & 22.57 $\pm$ 0.15 &                  & 24.27 $\pm$ 0.19 & 0.16 \\
EISD107 & 2 & 10.73 & 09 55 59.27 & -20 42 52.8 & 19.24 $\pm$ 0.02 & 21.05 $\pm$ 0.03 & 22.27 $\pm$ 0.05 & 0.03 \\
EISD110 & 3 &  N/A  & 09 54 53.46 & -21 15 13.1 & 20.01 $\pm$ 0.02 & 21.25 $\pm$ 0.03 & 23.94 $\pm$ 0.05 & 0.03 & N\\
EISD113 & 2 &  4.42 & 09 50 53.09 & -21 33 04.0 & 14.49 $\pm$ 0.00 & 14.39 $\pm$ 0.01 & 14.65 $\pm$ 0.01 & 0.98 \\
EISD116 & 1 & 72.10 & 09 51 29.71 & -20 16 42.8 & 20.13 $\pm$ 0.03 & 21.72 $\pm$ 0.08 & 22.77 $\pm$ 0.10 & 0.02 \\
EISD120 & 1 &  N/A  & 09 48 04.32 & -20 34 36.3 & 23.00 $\pm$ 0.30 &                  &                  & 0.00 & N\\
\hline
\end{tabular}
\end{table*}
\addtocounter{table}{-1}

\begin{table*}
\caption{{\it continued.} EIS optical counterparts of the radio sources.}
\begin{tabular}{lcrrrccccc}
\hline
\multicolumn{1}{c}{Source}  &
\multicolumn{1}{c}{Class}   &
\multicolumn{1}{c}{$L$}     &
\multicolumn{2}{c}{RA~~~~(J2000)~~~~Dec} &
\multicolumn{1}{c}{$I$-magnitude} &
\multicolumn{1}{c}{$V$-magnitude} &
\multicolumn{1}{c}{$B$-magnitude} &
\multicolumn{1}{c}{S/G} &
\multicolumn{1}{c}{Notes} \\
\hline
EISD122 & 1 & 43.12 & 09 52 55.92 & -20 51 45.0 & 18.68 $\pm$ 0.01 & 19.29 $\pm$ 0.01 & 19.41 $\pm$ 0.01 & 0.97 \\
EISD123 & 3 &  N/A  & \multicolumn{2}{c}{No position obtainable} & Saturated  & Saturated  & Saturated   & 0.00~ & N~\\
EISD126 & 2 &  1.94 & 09 54 36.24 & -21 44 30.9 & 22.54 $\pm$ 0.14 &                  &                  & 0.07 \\
EISD127 & 1 & 17.18 & 09 48 22.19 & -21 05 08.4 & 22.43 $\pm$ 0.12 &                  &                  & 0.00 \\
EISD132 & 2 &  5.74 & 09 46 19.15 & -20 37 57.7 & 17.41 $\pm$ 0.01 & 18.86 $\pm$ 0.01 & 19.84 $\pm$ 0.01 & 0.03 \\
EISD133 & 1 &  7.56 & 09 57 02.29 & -21 56 51.2 & 19.49 $\pm$ 0.03 & 22.22 $\pm$ 0.07 & 24.15 $\pm$ 0.14 & 0.03 \\
EISD134 & 2 &  0.68 & 09 46 49.47 & -21 16 46.4 & 18.36 $\pm$ 0.01 & 20.83 $\pm$ 0.02 & 21.82 $\pm$ 0.04 & 0.03 \\
EISD136 & 1 & 22.21 & 09 50 48.54 & -21 54 56.7 & 23.77 $\pm$ 0.25 & 24.69 $\pm$ 0.18 & 24.60 $\pm$ 0.22 & 0.65 \\
EISD137 & 1 & 11.63 & 09 48 14.19 & -19 59 56.3 & 18.93 $\pm$ 0.02 & 20.84 $\pm$ 0.02 & 21.92 $\pm$ 0.04 & 0.03 \\
EISD139 & 1 & 15.07 & 09 52 50.42 & -21 31 47.6 & 22.35 $\pm$ 0.14 & 24.33 $\pm$ 0.25 &                  & 0.41 \\
EISD141 & 1 & 10.38 & 09 55 11.46 & -20 30 19.2 & 17.52 $\pm$ 0.01 & 19.17 $\pm$ 0.01 & 20.25 $\pm$ 0.10 & 0.03 \\
EISD142 & 1 &  3.29 & 09 56 07.00 & -20 05 44.3 & 21.96 $\pm$ 0.10 & 24.37 $\pm$ 0.18 &                  & 0.02 \\
EISD143 & 1 & 78.31 & 09 57 35.36 & -20 29 35.4 & 19.97 $\pm$ 0.02 & 20.57 $\pm$ 0.03 & 20.90 $\pm$ 0.02 & 0.97 \\
EISD145 & 2 &  2.25 & 09 57 39.07 & -20 03 20.2 & 23.33 $\pm$ 0.19 &                  &                  & 0.74 \\
EISD146 & 1 & 51.27 & 09 56 42.30 & -21 19 44.3 & 23.50 $\pm$ 0.23 & 23.90 $\pm$ 0.17 & 22.95 $\pm$ 0.07 & 0.51 \\
EISD148 & 3 &  N/A  & 09 45 38.08 & -21 11 13.5 & 18.92 $\pm$ 0.02 & 21.38 $\pm$ 0.03 & 22.23 $\pm$ 0.05 & 0.03 & N\\
EISD149 & 1 & 71.30 & 09 47 44.76 & -21 12 23.4 & 18.34 $\pm$ 0.01 & 20.77 $\pm$ 0.02 & 21.73 $\pm$ 0.03 & 0.03 \\
EISD150 & 2 &  5.38 & 09 47 10.31 & -20 35 52.4 & 22.10 $\pm$ 0.11 & 24.30 $\pm$ 0.24 &                  & 0.07 \\
EISD153 & 1 & 48.44 & 09 56 49.79 & -20 35 25.9 & 17.09 $\pm$ 0.01 & 18.63 $\pm$ 0.01 & 19.70 $\pm$ 0.01 & 0.03 \\
EISD155 & 3 &  N/A  & 09 57 24.90 & -20 22 43.0 & 17.84 $\pm$ 0.01 & 20.10 $\pm$ 0.02 & 21.09 $\pm$ 0.02 & 0.03 & N\\
EISD156 & 1 & 49.46 & 09 56 37.10 & -20 19 05.8 & 16.94 $\pm$ 0.01 & 18.38 $\pm$ 0.01 & 19.60 $\pm$ 0.01 & 0.03 \\
EISD157 & 1 &  9.38 & 09 49 02.26 & -21 15 05.1 & 23.58 $\pm$ 0.26 &                  & 25.42 $\pm$ 0.19 & 0.48 \\
EISD159 & 1 &  9.01 & 09 53 57.43 & -20 36 51.0 & 21.10 $\pm$ 0.06 & 22.28 $\pm$ 0.06 & 22.41 $\pm$ 0.05 & 0.31 \\
EISD162 & 3 &  N/A  & 09 56 39.22 & -20 10 44.3 & 19.20 $\pm$ 0.30 &                  &                  & ---  & N\\
EISD163 & 3 &  N/A  & \multicolumn{2}{c}{No position obtainable} & Saturated  & Saturated  & Saturated   & 0.00~ & N~\\
EISD164 & 1 & 42.18 & 09 52 01.22 & -20 24 56.2 & 17.15 $\pm$ 0.01 & 18.58 $\pm$ 0.01 & 19.57 $\pm$ 0.01 & 0.03 \\
EISD165 & 2 & 11.66 & 09 54 10.51 & -21 58 01.5 & 22.73 $\pm$ 0.15 &                  &                  & 0.11 \\
EISD166 & 1 & 82.75 & 09 56 04.45 & -21 44 36.6 & 22.73 $\pm$ 0.22 & 23.62 $\pm$ 0.16 & 23.68 $\pm$ 0.10 & 0.48 \\
EISD169 & 2 &  1.31 & 09 51 49.02 & -21 33 40.0 & 17.15 $\pm$ 0.01 & 19.50 $\pm$ 0.01 & 19.97 $\pm$ 0.02 & 0.03 \\
EISD171 & 2 &  1.94 & 09 47 50.33 & -21 42 10.1 & 14.84 $\pm$ 0.00 & 14.99 $\pm$ 0.01 & 15.34 $\pm$ 0.01 & 0.98 & N\\
        & 2 &  0.41 & 09 47 50.64 & -21 42 11.4 & 17.43 $\pm$ 0.00 & 18.84 $\pm$ 0.01 & 19.34 $\pm$ 0.01 & 0.92 \\
EISD173 & 1 & 54.15 & 09 54 31.06 & -20 35 37.7 & 21.05 $\pm$ 0.07 & 22.96 $\pm$ 0.14 &                  & 0.00 \\
EISD174 & 1 & 28.36 & 09 49 02.79 & -20 16 11.0 & 22.50 $\pm$ 0.13 & 22.97 $\pm$ 0.12 & 23.99 $\pm$ 0.12 & 0.26 \\
EISD175 & 2 &  2.96 & 09 49 22.34 & -21 18 17.7 & 18.95 $\pm$ 0.02 & 21.83 $\pm$ 0.05 & 23.00 $\pm$ 0.25 & 0.03 \\
EISD177 & 1 & 80.27 & 09 55 26.95 & -20 46 05.9 & 19.99 $\pm$ 0.04 & 22.63 $\pm$ 0.06 & 23.45 $\pm$ 0.13 & 0.05 \\
EISD178 & 3 &  N/A  & 09 47 47.93 & -21 00 45.4 & 21.42 $\pm$ 0.08 & 22.63 $\pm$ 0.10 & 23.12 $\pm$ 0.06 & 0.52 & N\\
EISD179 & 1 & 10.44 & 09 49 59.71 & -21 27 18.3 & 19.95 $\pm$ 0.03 & 22.29 $\pm$ 0.07 & 22.60 $\pm$ 0.15 & 0.04 \\
EISD180 & 1 & 48.63 & 09 49 12.74 & -22 00 23.4 & 17.61 $\pm$ 0.01 & 19.85 $\pm$ 0.02 & 20.81 $\pm$ 0.02 & 0.03 \\
EISD181 & 1 & 20.07 & 09 54 41.88 & -20 49 43.4 & 24.85 $\pm$ 0.35 &                  &                  & 0.37 & N\\
EISD185 & 1 & 14.39 & 09 52 14.36 & -21 40 18.4 & 14.84 $\pm$ 0.00 & 15.87 $\pm$ 0.01 & 16.15 $\pm$ 0.01 & 0.04 \\
EISD186 & 1 & 35.36 & 09 49 24.62 & -21 11 11.6 & 20.77 $\pm$ 0.04 & 22.83 $\pm$ 0.07 & 23.10 $\pm$ 0.10 & 0.04 \\
EISD187 & 2 &  0.42 & 09 50 38.66 & -21 41 11.4 & 18.73 $\pm$ 0.02 & 21.49 $\pm$ 0.05 & 22.43 $\pm$ 0.08 & 0.03 \\
EISD191 & 3 &  N/A  & \multicolumn{2}{c}{No position obtainable} & Saturated  & Saturated  & Saturated   & 0.00~ & N~\\
EISD199 & 1 & 53.54 & 09 45 26.32 & -21 55 00.1 & 17.99 $\pm$ 0.01 & 19.64 $\pm$ 0.01 & 20.71 $\pm$ 0.02 & 0.03 \\
\hline
\end{tabular}
\end{table*}

\section{The final CENSORS sample}
\label{appfinalsamp}

Table~\ref{finalsamptab} provides the definition of the final CENSORS sample.

\begin{table*}
\caption{\label{finalsamptab} Definition of the final CENSORS sample. This
includes only those NVSS sources which are real and fall within the EIS Patch
D. It is numbered in order of decreasing flux density, as determined from the
latest NVSS catalogue (version 2.17), although those sources marked with an
asterisk have had their NVSS flux densities adjusted to account for overlap
with another NVSS source. Four additional sample members are appended, as
discussed in the text.}
\begin{tabular}{llccllccllc}

\multicolumn{1}{c}{New} & 
\multicolumn{1}{c}{Old~~~} & 
$S_{\rm 1.4\,GHz}^{\rm NVSS}$ & \hspace*{0.3cm} & 
\multicolumn{1}{c}{New} & 
\multicolumn{1}{c}{Old~~~} & 
$S_{\rm 1.4\,GHz}^{\rm NVSS}$ & \hspace*{0.3cm} &
\multicolumn{1}{c}{New} & 
\multicolumn{1}{c}{Old~~~~~~} & 
$S_{\rm 1.4\,GHz}^{\rm NVSS}$ \\

\multicolumn{1}{c}{name} & 
\multicolumn{1}{c}{name~~~} & 
(mJy) &&
\multicolumn{1}{c}{name} & 
\multicolumn{1}{c}{name~~~} & 
(mJy) &&
\multicolumn{1}{c}{name} & 
\multicolumn{1}{c}{name~~~~~~} & 
(mJy) \\

CENSORS1   & EISD1    & 659.5 $\pm$ 19.8  && CENSORS53  & EISD76   &  21.6 $\pm$  1.1  && CENSORS105 & EISD138  &  10.6 $\pm$  0.6  \\
CENSORS2   & EISD2    & 452.3 $\pm$ 13.6  && CENSORS54  & EISD74   &  21.4 $\pm$  0.8  && CENSORS106 & EISD142  &  10.5 $\pm$  0.6  \\
CENSORS3   & EISD3    & 355.3 $\pm$ 10.7  && CENSORS55  & EISD71   &  21.4 $\pm$  0.8  && CENSORS107 & EISD148  &  10.3 $\pm$  1.0  \\
CENSORS4   & EISD6    & 283.0 $\pm$  9.5  && CENSORS56  & EISD78   &  20.8 $\pm$  1.1  && CENSORS108 & EISD153  &  10.2 $\pm$  0.6  \\
CENSORS5   & EISD8    & 244.7 $\pm$  8.2  && CENSORS57  & EISD80   &  20.7 $\pm$  1.1  && CENSORS109 & EISD154  &  10.1 $\pm$  0.6  \\
CENSORS6   & EISD7*   & 239.7 $\pm$  1.3  && CENSORS58  & EISD79   &  20.7 $\pm$  0.8  && CENSORS110 & EISD141  &  10.1 $\pm$  1.3  \\
CENSORS7   & EISD10   & 148.2 $\pm$  5.1  && CENSORS59  & EISD81   &  19.1 $\pm$  1.1  && CENSORS111 & EISD149  &  10.0 $\pm$  0.6  \\
CENSORS8   & EISD11   & 126.3 $\pm$  3.8  && CENSORS60  & EISD83   &  18.9 $\pm$  0.7  && CENSORS112 & EISD146  &   9.8 $\pm$  0.6  \\
CENSORS9   & EISD12   & 118.2 $\pm$  3.6  && CENSORS61  & EISD82   &  18.5 $\pm$  0.7  && CENSORS113 & EISD150  &   9.7 $\pm$  0.6  \\
CENSORS10  & EISD16   &  79.4 $\pm$  2.9  && CENSORS62  & EISD84   &  18.4 $\pm$  0.7  && CENSORS114 & EISD166  &   9.6 $\pm$  0.6  \\
CENSORS11  & EISD15   &  78.1 $\pm$  2.4  && CENSORS63  & EISD88   &  18.3 $\pm$  0.7  && CENSORS115 & EISD155  &   9.6 $\pm$  1.0  \\
CENSORS12  & EISD18   &  70.4 $\pm$  2.6  && CENSORS64  & EISD85   &  18.1 $\pm$  1.0  && CENSORS116 & EISD143  &   9.6 $\pm$  0.6  \\
CENSORS13  & EISD20   &  66.3 $\pm$  2.7  && CENSORS65  & EISD87   &  17.9 $\pm$  1.0  && CENSORS117 & EISD165  &   9.5 $\pm$  0.6  \\
CENSORS14  & EISD21   &  65.6 $\pm$  2.4  && CENSORS66  & EISD90   &  17.4 $\pm$  1.1  && CENSORS118 & EISD161  &   9.4 $\pm$  0.6  \\
CENSORS15  & EISD22   &  63.0 $\pm$  1.9  && CENSORS67  & EISD89   &  17.3 $\pm$  0.7  && CENSORS119 & EISD157  &   9.4 $\pm$  0.6  \\
CENSORS16  & EISD23   &  61.7 $\pm$  2.3  && CENSORS68  & EISD91   &  17.2 $\pm$  0.7  && CENSORS120 & EISD159  &   9.1 $\pm$  0.6  \\
CENSORS17  & EISD24   &  61.5 $\pm$  2.3  && CENSORS69  & EISD92   &  17.0 $\pm$  0.7  && CENSORS121 & EISD164  &   9.0 $\pm$  0.5  \\
CENSORS18  & EISD25   &  58.3 $\pm$  1.8  && CENSORS70  & EISD124* &  17.0 $\pm$  2.0  && CENSORS122 & EISD156  &   9.0 $\pm$  0.6  \\
CENSORS19  & EISD27   &  55.1 $\pm$  2.1  && CENSORS71  & EISD93   &  16.7 $\pm$  0.7  && CENSORS123 & EISD173  &   8.7 $\pm$  0.5  \\
CENSORS20  & EISD30   &  54.2 $\pm$  2.1  && CENSORS72  & EISD97   &  16.5 $\pm$  0.7  && CENSORS124 & EISD163  &   8.7 $\pm$  0.6  \\
CENSORS21  & EISD28   &  54.0 $\pm$  1.7  && CENSORS73  & EISD94   &  16.2 $\pm$  0.7  && CENSORS125 & EISD175  &   8.4 $\pm$  0.5  \\
CENSORS22  & EISD29   &  52.9 $\pm$  1.7  && CENSORS74  & EISD96   &  16.0 $\pm$  0.7  && CENSORS126 & EISD171  &   8.4 $\pm$  1.3  \\
CENSORS23  & EISD31   &  52.4 $\pm$  2.0  && CENSORS75  & EISD98   &  15.7 $\pm$  1.0  && CENSORS127 & EISD186  &   8.3 $\pm$  0.5  \\
CENSORS24  & EISD32   &  51.0 $\pm$  1.6  && CENSORS76  & EISD102  &  15.3 $\pm$  0.7  && CENSORS128 & EISD174  &   8.3 $\pm$  0.5  \\
CENSORS25  & EISD34   &  49.2 $\pm$  1.9  && CENSORS77  & EISD104  &  15.0 $\pm$  0.7  && CENSORS129 & EISD170  &   8.3 $\pm$  0.6  \\
CENSORS26  & EISD36   &  44.4 $\pm$  1.4  && CENSORS78  & EISD107  &  14.6 $\pm$  0.7  && CENSORS130 & EISD172  &   8.2 $\pm$  0.5  \\
CENSORS27  & EISD44*  &  40.4 $\pm$  2.3  && CENSORS79  & EISD106  &  14.6 $\pm$  1.1  && CENSORS131 & EISD169  &   8.2 $\pm$  0.6  \\
CENSORS28  & EISD38   &  40.1 $\pm$  1.9  && CENSORS80  & EISD110  &  14.5 $\pm$  0.6  && CENSORS132 & EISD167  &   7.9 $\pm$  0.6  \\
CENSORS29  & EISD39   &  38.2 $\pm$  1.6  && CENSORS81  & EISD105  &  14.5 $\pm$  1.4  && CENSORS133 & EISD183  &   7.8 $\pm$  1.2  \\
CENSORS30  & EISD40   &  37.8 $\pm$  2.0  && CENSORS82  & EISD113  &  13.6 $\pm$  0.6  && CENSORS134 & EISD182  &   7.8 $\pm$  0.6  \\
CENSORS31  & EISD41   &  37.3 $\pm$  1.5  && CENSORS83  & EISD116  &  13.5 $\pm$  0.6  && CENSORS135 & EISD178  &   7.8 $\pm$  0.6  \\
CENSORS32  & EISD43   &  35.3 $\pm$  1.5  && CENSORS84  & EISD103  &  13.5 $\pm$  1.2  && CENSORS136 & EISD181  &   7.5 $\pm$  0.6  \\
CENSORS33  & EISD45   &  34.3 $\pm$  1.1  && CENSORS85  & EISD112  &  13.4 $\pm$  1.0  && CENSORS137 & EISD187  &   7.4 $\pm$  1.2  \\
CENSORS34  & EISD47   &  34.2 $\pm$  1.1  && CENSORS86  & EISD120  &  13.2 $\pm$  0.6  && CENSORS138 & EISD177  &   7.1 $\pm$  0.6  \\
CENSORS35  & EISD48   &  34.1 $\pm$  1.4  && CENSORS87  & EISD111  &  13.2 $\pm$  0.6  && CENSORS139 & EISD180  &   6.9 $\pm$  0.6  \\
CENSORS36  & EISD51   &  32.3 $\pm$  1.1  && CENSORS88  & EISD119  &  13.1 $\pm$  0.6  && CENSORS140 & EISD199  &   6.8 $\pm$  0.5  \\
CENSORS37  & EISD52   &  31.8 $\pm$  1.4  && CENSORS89  & EISD117  &  13.0 $\pm$  1.0  && CENSORS141 & EISD189  &   6.6 $\pm$  0.6  \\
CENSORS38  & EISD53   &  31.7 $\pm$  1.1  && CENSORS90  & EISD114  &  12.8 $\pm$  0.6  && CENSORS142 & EISD195  &   6.3 $\pm$  0.6  \\
CENSORS39  & EISD54   &  31.5 $\pm$  1.1  && CENSORS91  & EISD127  &  12.7 $\pm$  0.6  && CENSORS143 & EISD188  &   6.1 $\pm$  0.6  \\
CENSORS40  & EISD55   &  30.9 $\pm$  1.3  && CENSORS92  & EISD122  &  12.6 $\pm$  1.1  && CENSORS144 & EISD179  &   6.0 $\pm$  0.6  \\
CENSORS41  & EISD58   &  27.5 $\pm$  1.7  && CENSORS93  & EISD132  &  12.2 $\pm$  0.6  && CENSORS145 & EISD137* &   5.8 $\pm$  0.3  \\
CENSORS42  & EISD60   &  26.5 $\pm$  0.9  && CENSORS94  & EISD125  &  12.2 $\pm$  0.6  && CENSORS146 & EISD191  &   5.4 $\pm$  0.6  \\
CENSORS43  & EISD64   &  26.4 $\pm$  0.9  && CENSORS95  & EISD123  &  12.2 $\pm$  1.2  && CENSORS147 & EISD197  &   4.2 $\pm$  0.7  \\
CENSORS44  & EISD62   &  26.1 $\pm$  0.9  && CENSORS96  & EISD131  &  12.0 $\pm$  0.6  && CENSORS148 & EISD162  &   4.1 $\pm$  0.8  \\
CENSORS45  & EISD66   &  25.5 $\pm$  1.2  && CENSORS97  & EISD126  &  12.0 $\pm$  1.2  && CENSORS149 & EISD185  &   4.0 $\pm$  0.7  \\
CENSORS46  & EISD65   &  25.2 $\pm$  0.9  && CENSORS98  & EISD130  &  11.8 $\pm$  0.6  && CENSORS150 & EISD194  &   3.8 $\pm$  0.7  \\
CENSORS47  & EISD63   &  25.2 $\pm$  0.9  && CENSORS99  & EISD133  &  11.6 $\pm$  0.6  && CENSORS-X1 &J094651-2125 &7.2 $\pm$  0.5  \\
CENSORS48  & EISD68   &  24.2 $\pm$  0.9  && CENSORS100 & EISD136  &  11.5 $\pm$  0.6  && CENSORS-X2 &J095233-2129 &6.8 $\pm$  0.5  \\
CENSORS49  & EISD67   &  23.8 $\pm$  0.9  && CENSORS101 & EISD139  &  11.4 $\pm$  0.6  && CENSORS-X3 &J095240-2123 &6.7 $\pm$  0.5  \\
CENSORS50  & EISD69   &  22.3 $\pm$  0.8  && CENSORS102 & EISD134  &  11.1 $\pm$  1.1  && CENSORS-D1 &J095218-2038 &3.6 $\pm$  0.6  \\
CENSORS51  & EISD75   &  21.7 $\pm$  0.8  && CENSORS103 & EISD56*  &  10.7 $\pm$  0.6  &&            &J095223-2041 &3.8 $\pm$  0.7  \\
CENSORS52  & EISD72   &  21.7 $\pm$  0.8  && CENSORS104 & EISD145  &  10.7 $\pm$  0.6  \\
\end{tabular}
\end{table*}

\label{lastpage}
\end{document}